\documentclass[twocolumn, showpacs, prb]{revtex4}
\usepackage{hyperref}
\usepackage{amsmath}
\usepackage{amssymb}
\usepackage{graphicx}
\usepackage{multirow}
\begin{document}
\title{Reentrant behavior of superconducting alloys}
\date{\today}
\author{Dawid Borycki}
\email{dawid.borycki@fizyka.umk.pl}
\affiliation{Instytut Fizyki, Uniwersytet M. Kopernika, ul. Grudziadzka 5, 87-100 Torun, Poland}
\author{Jan Ma\'{c}kowiak}
\email{ferm92@fizyka.umk.pl}
\affiliation{Instytut Fizyki, Uniwersytet M. Kopernika, ul. Grudziadzka 5, 87-100 Torun, Poland}

\begin{abstract}
A dirty BCS superconductor with magnetic impurities is studied. Asymptotic solution of the thermodynamics of such superconductor with spin $1/2$ and $7/2$ magnetic impurities, is found. To this end, the system's free energy $f(H, \beta)$ is bounded from above and below by mean-field type bounds, which are shown to coalesce almost exactly in the thermodynamic limit, provided the impurity concentration is sufficiently small. The resulting mean-field equations for the gap $\Delta$ and a parameter $\nu$, characterizing the impurity subsystem, are solved and the solution minimizing $f$ is found for various values of magnetic coupling constant $g$ and impurity concentration $x$. The phase diagrams of the system are depicted with five distinct phases: the normal phase, unperturbed superconducting phase, perturbed
superconducting phase with nonzero gap in the excitation spectrum, perturbed gapless superconducting phase and impurity phase with completely suppressed superconductivity. Furthermore, evidence of reentrant superconductivity and Jaccarino-Peter compensation is found. The credibility of the theory is verified by testing the dependence of the superconducting transition temperature $T_{\text{c}}$ on $x$. Very good quantitative agreement with experimental data is obtained for several alloys: (La$_{1-x}$Ce$_{x}$)Al$_{2}$, (La$_{1-x}$Gd$_{x}$)Al$_{2}$ and (La$_{0.8-x}$Y$_{0.20}$)Ce$_{x}$. The theory presented improves earlier developments in this field.

\end{abstract}

\maketitle

\section{Introduction}
It is well known that doping may substantially change the properties of a superconductor. The superconducting transition temperature $T_{\text{c}}$ in most dirty superconductors decreases with impurity concentration $c$ \cite{Matthias1}. However, alloys of zirconium with iron, cobalt or nickel have higher transition temperatures than pure zirconium \cite{Matthias2}. Similar behavior was observed in titanium doped by chromium, manganese, iron and cobalt \cite{Matthias3}. One of the properties, which do not change under doping, is the order of the transition in the absence of external magnetic field \cite{Muller, LaGd}. 

Early attempts to explain lowered $T_{\mathrm{c}}$ in the presence of magnetic impurities were founded on perturbation theory. Nakamura \cite{Nakamura} and Suhl et al. \cite{Suhl} explained this effect by treating the s-d interaction $V_{\mathrm{s-d}}$ \cite{Kasuya} as an additive term in the total Hamiltonian, which perturbs a BCS superconductor \cite{BCS}. However, their theory predicts a first order phase transition to the superconducting state in zero magnetic field. 

The well known Abrikosov-Gor'kov theory \cite{AG} (AG) of dirty superconductors explains the strong decrease in $T_{\mathrm{c}}$ due to magnetic impurities and also predicts "gapless superconductivity", confirmed experimentally by Reif and Woolf \cite{ReifWoolf}. Disagreement with this approach is observed in a number of Kondo superconductors, e.g. La$_{1-x}$Ce$_{x}$Al$_{2}$ \cite{MapleConvAndUnconvSC}, LaCe and LaGd \cite{ChaikinMihalisin} and PbCe and InCe \cite{DelfsBeaudryFinnemore}. 

Such compunds manifest reentrant superconductivity (RSC) which is due to competition between the Kondo effect and superconductivity. The RSC effect was predicted theoretically by M\"{u}ller-Hartmann and Zittartz \cite{MHZ} (MHZ) and is expected to occur when the superconducting transition temperature of the host compound $T_{\text{c}0}$ is much larger than the Kondo temperature $T_{\text{K}}$. In such the case the equation for $T_{\text{c}}(c)$ has three solutions: $T_{\text{c}1}$, $T_{\text{c}2}$ and $T_{\text{c}3}$ for certain values of impurity concentration. As the temperature $T$ is lowered the alloy becomes superconducting at $T_{\text{c}1}$. When the value of $T$ is comparable to $T_{\text{K}}$ the pair-breaking effect of impurities increases and superconductivity is suppressed at $T_{\text{c}2}$. The alloy re-enters the superconducting state at $T_{\text{c}3}$, where the pair-breaking passes through its maximum value. Such reentrant behavior was observed experimentally in La$_{0.7915}$Ce$_{0.0085}$Y$_{0.20}$ with $T_{\text{c}1} = 0.55\,\text{K},\ T_{\text{c}2} = 0.27\,\text{K}$ and $T_{\text{c3}} = 0.05 \,\text{K}$\cite{Winzer}. 

According to M\"{u}ller-Hartmann and Zittartz theory, superconductivity is never completely suppressed for any value of impurity concentration if $T_{\text{K}}/T_{\text{c}0} \ll 1$. This statement disagrees with experiment, which shows that disappearance of superconductivity above a critical value of $c$ is possible. Furthermore, significant deviations from this approach were observed for (La, Ce)In$_{3-x}$Sn$_{x}$\cite{MaarenVanHaeringen}. 

First experimental observations of reentrant superconductivity \cite{MapleRSC, RibletRSC} revealed no evidence of $T_{\text{c}3}$, corresponding to second phase transition, reintroducing superconductivity. For this reason, MHZ theory was reformulated (e.g.~Refs.~\onlinecite{PS75a, PS75b, TM76, MHZ76}) and also other proposals for a theory of $T_{\text{c}}(c)$ were given. In particular Jarrell, performed Monte Carlo simulations of the superconducting transition temperature in terms of Eliashberg-Migdal perturbation theory \cite{JM99}. These calculations raised doubts about the existence of $T_{\text{c}3}$, contrary to experiments accomplished by Winzer \cite{Winzer}, confirming the presence of a transition back to the superconducting state.      

The effect of magnetic impurities on superconductivity is still under debate. Recently, Barzykin and Gor'kov \cite{BG05} studied s-wave superconductivity in the Anderson lattice and demonstrated excellent agreement of the resulting  $T_{\text{c}}(x)$ graphs for Ce$_{1-x}$La$_{x}$Ru$_{3}$Si$_{2}$ with experiment. Reentrant behavior of $T_{\text{c}}$ may occur for untypical values of parameters. Kozorezov et al.\cite{AK08} have shown, in terms of the MHZ model, that trace concentrations of magnetic impurities may also result in significant changes in nonequilibrium properties of superconductors. A~comprehensive review of recent developments in this field can be found in Ref.~\onlinecite{AB06}.
 
Experimental studies of superconductors containing magnetic impurities carried out by Matthias \cite{Matthias1} revealed another extraordinary property, viz., the coexistence of magnetism and superconductivity. Till then these two phenomena were believed to be mutually exclusive, since the internal magnetic fields generated in magnetically ordered systems are much larger than the typical critical fields of superconductors. 
The coexistence hypothesis was confirmed shortly after in the following superconducting alloys: Ce$_{1-x}$Gd$_{x}$Ru$_{2}$~\cite{MatthiasFerromagneticSC} and~Y$_{1-x}$Gd$_{x}$Os$_{2}$\cite{HS59}, although it was not found to occur in the same volume element. The phase diagrams obtained by Wilhelm and Hillenbrand for Ce$_{1-x}$Tb$_{x}$Ru$_{2}$\cite{MW70} also contains the coexistence phase. Specific-heat measurements showed short-range ordering in this alloy, typical of spin-glass systems\cite{MP71}.

The coexistence of superconductivity and long-range antiferromagnetic ordering of the rare earth R magnetic moments was discovered in RMo$_{6}$Se$_{8}$ (R = Gd, Tb and Er)\cite{McCallumASC}, RRh$_{4}$B$_{4}$ (R = Nd, Sm and Tm)\cite{HamakerASC} and in RMo$_{6}$S$_{8}$ (R = Gd, Tb, Dy and Er)\cite{IshikawaASC}. A similar overlap between superconductivity and ferromagnetism was observed in ErRh$_{4}$B$_{4}$\cite{MapleFSC} and HoMo$_{6}$S$_{8}$ \cite{IshikawaFSC}. 

The phase diagrams of superconducting alloys, containing the coexistence phase, have been computed by several theoretists. Gor'kov and Rusinov extended the AG theory to include magnetic ordering. Correspondence with the phase diagrams observed experimentally is expected to occur for very strong spin-orbit scattering. Balseiro et al. \cite{Balseiro} studied a BCS superconductor perturbed by magnetic impurities interacting via a nearest neighbour Heisenberg potential. The resulting phase diagrams comply qualitatively with experiment. 

Theories of dirty superconductors contribute significantly to our understanding of the superconductivity
phenomenon. One can expect that further investigations of superconducting alloys will explain the microscopic mechanism of unconventional
superconductivity displayed by some materials, e.g. high
temperature superconductors\cite{BM} and heavy fermion compounds \cite{AdresGrebnerOtt}.
 
This issue, as well as some shortcomings of the models presented above, motivate the present work. We investigate a BCS Hamiltonian
\cite{BCS}
\begin{equation}
H_{\text{BCS}} = H_{0} + V_{\text{BCS}},
\end{equation}
supplemented by a reduced s-d interaction
\begin{equation}
\label{SDreduced}
V = g^{2} N^{-1} \sum_{\alpha=1}^{M} \sigma_{z} S_{z\alpha}, 
\end{equation}
where
\[
H_{0} = \sum_{\mathbf{k}\sigma} \xi_{\mathbf{k}} n_{\mathbf{k}\sigma},
\] 
with $\xi_{\mathbf{k}} = \varepsilon_{\mathbf{k}} - \mu$, $n_{\mathbf{k}\sigma} = a_{\mathbf{k}\sigma}^{\dagger}a_{\mathbf{k}\sigma}^{}$, is the free fermion kinetic energy operator and
\begin{equation}
\label{VBCS}
V_{\text{BCS}} = - |\Lambda|^{-1} \sum_{\mathbf{k} \mathbf{k^{\prime}}} G_{\mathbf{k}\, \mathbf{k^{\prime}}} a_{\mathbf{k}+}^{\dagger} a_{-\mathbf{k}-}^{\dagger} a_{-\mathbf{k}^{\prime}-}^{} a_{\mathbf{k}^{\prime}+},
\end{equation}
is the Cooper pairing potential. $|\Lambda|$ denotes the system's volume and $G_{\mathbf{k}\, \mathbf{k}'}^{}$ is real, symmetric, invariant under $\mathbf{k}\rightarrow - \mathbf{k}$ or $\mathbf{k}'\rightarrow - \mathbf{k}'$ and nonvanishing only in a thin band close to the Fermi surface, viz.,
\[
G_{\mathbf{k}\, \mathbf{k}'}^{} = G_{0}\chi(\mathbf{k})\chi(\mathbf{k}'), \qquad G_{0} > 0,
\]
where $\chi(\mathbf{k})$ denotes the characteristic function of the set
\[
{\cal P} = \{\mathbf{k}: \mu - \delta \leq \varepsilon_{\mathbf{k}}^{} \leq \mu + \delta \}, \qquad \varepsilon_{\mathbf{k}}^{} = \frac{\hbar^{2} \mathbf{k}^{2}}{2m}.
\]
In equation (\ref{SDreduced}), $S_{z\alpha}$ denotes the spin operator of the magnetic ion, whereas 
\[
\sigma_{z} = \sum_{\mathbf{k}}\left(n_{\mathbf{k+}} - n_{\mathbf{k-}}\right),
\]
describes the spin operator of each conducting fermion. $M$ is the number of magnetic impurities, $N$ the number of host atoms.

We assume the perturbation implemented by the localized distinguishable magnetic impurities to be a reduced long-range s-d interaction, which involves only the z-components of the impurity and fermion spin operators (Eq.~(\ref{SDreduced})). The reason for this simplification is that the thermodynamics of the resulting Hamiltonian $H=H_{0} + V_{\text{BCS}} + V$ admits a mean-field solution which improves with decreasing impurity density. Furthermore, this solution is thermodynamically equivalent to the one obtained for $H$ with a Heisenberg type reduced s-d interaction
\begin{equation}
\begin{split}
V_{\text{H}}^{} & = -\frac{g^2}{N} \sum_{\mathbf{k}\alpha} \Bigl[(a_{\mathbf{k}-}^{\dagger} a_{\mathbf{k}-}^{} - a_{\mathbf{k}+}^{\dagger} a_{\mathbf{k}+}^{}) S_{z\alpha}^{} \\& -  a_{\mathbf{k}+}^{\dagger} a_{\mathbf{k}-}^{}S_{\alpha -}^{} - a_{\mathbf{k}-}^{\dagger} a_{\mathbf{k}+}^{}S_{\alpha +}^{}\Bigr], 
\end{split}
\end{equation}
\[
S_{\alpha \pm}^{} = S_{\alpha x}^{} \pm \text{i} S_{\alpha y}^{},
\]
replacing $V$ (Ref.~\onlinecite{JM99}, Sec. 6.2.5). (Similarly, the thermodynamics of a classical superconductor can be explained in terms of a reduced BCS interaction, whereas a gauge-invariant theory of the Meissner effect requires a more general pairing potential.) The reduced form of $V_{H}$, obtained by rejecting the sum $\sum\limits_{\mathbf{k} \neq \mathbf{k}^{\prime}} V_{\mathbf{k} \, \mathbf{k}^{\prime}}$ in the s-d interaction $V_{\text{s-d}}$, is an approximation which resorts to the fact that spin-exchange processes, and not momentum exchange processes, are primarily responsible for the Kondo effect caused by $V_{\text{s-d}}$ \cite{Kondo}.

In Sections~\ref{Sec:LowerBound}~and~\ref{Sec:UpperBound} the upper and lower bound to the system's free energy $f(H,\beta)$ are derived by exploiting the method developed in Refs.~\onlinecite{MackowiakPA, TC1, TC2, Rickayzen}. These bounds are shown to be almost equal if the impurity density $M/|\Lambda|$ is sufficiently small (Sec.~\ref{Sec:BoundsEquality}). The mean-field equations for the gap $\Delta$ and parameter $\nu$, characterizing the impurity subsystem are derived in Secs.~\ref{Sec:FreeEnergyMinimization}--\ref{Sec:ChemicalPotential} and solved in Sec.~\ref{Sec:PhaseDiagrams} for various values of $g$, $M$, $G_{0}$ and $\delta$. In section~\ref{Sec:CriticalTemperature} the values of these parameters are adjusted to fit the experimental $T_{\text{c}}(x)$ curves for La$_{1-x}$Ce$_{x}$Al$_{2}$, La$_{1-x}$Gd$_{x}$Al$_{2}$ and La$_{0.8-x}$Y$_{0.20}$Ce$_{x}$. Quantitative agreement is found for each alloy. Furthermore, the phase diagrams derived in Sec.~\ref{Sec:PhaseDiagrams} qualitatively reproduce the experimental observations of the coexistence phase, reentrant behavior, gapless superconductivity and Jaccarino-Peter compensation.

\section{Lower bound to the free energy}
\label{Sec:LowerBound}
A version of the Tindemans and Capel method\cite{TC1, TC2}, introduced recently in order to study thermodynamics of the Fermi gas interacting with randomly distributed magnetic impurities\cite{MackowiakPA}, will be applied in this section to derive a lower bound to the system's free energy.

The full Hamiltonian of the system is
\begin{equation}
\label{Hsd}
H^{(M)} = T_{0} + V_{\text{BCS}} + V.
\end{equation}
Exploiting the identity
\[
g^{2}\sigma_{z} S_{z\alpha} = -\tfrac{1}{2}g^{2}\left(\sigma_{z} -  S_{z\alpha} \right)^{2} + \tfrac{1}{2} g^{2} \sigma_{z}^{2} + \tfrac{1}{2} g^{2} S_{z\alpha}^{2},
\]
the partition function can be written in the following form
\begin{equation}
\label{Z}
\begin{split}
Z(M) & = \text{Tr} \exp(-\beta H^{(M)}) = \text{Tr} \exp \Biggl[-\beta H_{\text{BCS}} \\& + \frac{g^{2} \beta}{2N} \sum_{\alpha=1}^{M} \Bigl(\bigl( \sigma_{z} - S_{z\alpha} \bigr)^{2} - \sigma_{z}^{2} - S_{z\alpha}^{2}\Bigr) \Biggr].
\end{split}
\end{equation}
The Gaussian integral 
\[
\exp\left(a^{2}\right) = \frac{1}{\sqrt{2 \pi}} \int_{-\infty}^{\infty} \exp \left( -\tfrac{1}{2} \nu^{2} + \sqrt{2}a\nu\right)\text{d}\nu
\]
and the commutation relations
\begin{equation}
\label{CRbcs}
\left[V_{\text{BCS}},\ \sum_{\mathbf{k}^{\prime\prime}}n_{\mathbf{k}^{\prime\prime}-}\right] = \left[V_{\text{BCS}},\  \sum_{\mathbf{k}^{\prime\prime}}n_{\mathbf{k}^{\prime\prime}+}\right] = 0,
\end{equation}
satisfied for $G_{\mathbf{k}\,\mathbf{k}^{\prime}}$ of the form $G_{\mathbf{k}\,\mathbf{k}^{\prime}} = G_{0} \chi(\mathbf{k}) \chi(\mathbf{k^{\prime}})$, allow to separate the electron and impurity spin operators in (\ref{Z}), 
\begin{equation}
\begin{split}
Z(M) & = \left( \frac{\beta N}{2\pi} \right)^{M/2} \text{Tr} \Biggl\{ \exp[-\beta H_{\text{BCS}}] \\& \times \int_{-\infty}^{\infty}\prod_{\alpha=1}^{M}\text{d}\nu_{\alpha} \exp\Bigl[ -\tfrac{1}{2} \beta \nu_{\alpha}^{2} + \beta g \nu_{\alpha} \left( \sigma_{z} - S_{z\alpha}\right)\\& - \tfrac{1}{2} \beta N^{-1}  g^{2} \sigma_{z}^{2} - \tfrac{1}{2} \beta N^{-1} g^{2} S_{z\alpha}^{2}\Bigr]\Biggr\},
\end{split}
\end{equation}
where $H_{\text{BCS}} = H_{0} + V_{\text{BCS}}$. In order to linearize quadratic terms in $\sigma_{z}$ in the exponent of the integrand on the r.h.s. one exploits the inequality
\[
\begin{split}
\tfrac{1}{2} N^{-1} g^{2} \sigma_{z}^{2} & = \tfrac{1}{2} N^{-1} \left( g \sigma_{z} - N \eta \right)^{2} + g \eta \sigma_{z} - \tfrac{1}{2} N \eta^{2} \\& \geq g \eta \sigma_{z} - \tfrac{1}{2} N \eta^{2},
\end{split}
\]
where $\eta$ is an arbitrary function of $\nu_{1}, \dots, \nu_{M}$. This yields an upper bound to the partition function
\begin{equation}
\begin{split}
Z(M) \leq& \left( \frac{\beta N}{2\pi} \right)^{M/2} \text{Tr} \exp \left[-\beta H_{\text{BCS}} \right] \\& \times \int_{-\infty}^{\infty} \prod_{\alpha=1}^{M} \text{d}\nu_{\alpha} \exp \Bigl[ -\tfrac{1}{2} N^{-1} g^{2} \beta S_{z\alpha}^{2} - g \beta \nu_{\alpha} S_{z\alpha} \\&- \tfrac{1}{2} N \beta \bigl( \nu_{\alpha}^{2} - \eta^{2} \bigr) + \beta g \bigl( \nu_{\alpha} - \eta\bigr) \sigma_{z}\Bigr].
\end{split}
\end{equation}
Following Pearce and Thompson~\cite{Pearce}, let us now add and subtract the term $\tfrac{1}{2} N \beta x^{-1}\nu_{\alpha}^{2},$ with $x>1$ in the exponent of the integrand. The resulting expression is next split into two factors and one of them is replaced by its maximum with respect to $\{\nu_{\alpha} \}$:
\begin{equation}
\label{Zbl}
\begin{split}
Z(M) &\leq \left( \frac{\beta N}{2\pi} \right)^{M/2} \max_{\{\nu_{\alpha}\}}\exp \bigl(G(\nu_{1}, \dots, \nu_{M}, \eta)\bigr) \\& \times\int_{-\infty}^{\infty} \prod_{\alpha^{\prime}}^{M}\text{d}\nu_{\alpha^{\prime}} \exp \left[ -\tfrac{1}{2} N \beta \nu_{\alpha^{\prime}}^{2} \bigl( 1 - x^{-1}\bigr)\right] \\
& = \max_{\{\nu_{\alpha}\}}\exp \bigl(G(\nu_{1}, \dots, \nu_{M}, \eta)\bigr)\left(1 - x^{-1}\right)^{-M/2},
\end{split}
\end{equation}
where
\begin{equation}
\begin{split}
\label{G}
G(\nu_{1}, &\dots, \nu_{M}, \eta)  = \ln \text{Tr} \exp \Biggl[ -\beta H_{\text{BCS}} \\& + \beta \sum_{\alpha = 1}^{M} \Bigl[g \bigl( \nu_{\alpha} - \eta\bigr) \sigma_{z} -\tfrac{1}{2}N x^{-1}\nu_{\alpha}^{2} \\& - \tfrac{1}{2} N^{-1} g^{2} S_{z\alpha}^{2} - g \nu_{\alpha} S_{z\alpha} + \tfrac{1}{2} N \eta^{2}\Bigr]\Biggr].
\end{split}
\end{equation}
Inequality (\ref{Zbl}) yields the relevant lower bound to the free energy $|\Lambda|f(H,\beta) = -\beta^{-1} \ln Z$,
\begin{equation}
\label{fLowerBound}
\begin{split}
|\Lambda| f(H,\beta) \geq &- \min_{\{\nu_{\alpha}\}} \beta^{-1} G(\nu_{1}, \dots, \nu_{M}, \eta) \\& + \tfrac{1}{2}\beta^{-1} M \ln \left(1 - x^{-1}\right).
\end{split}
\end{equation}
The function $\eta$ will be now chosen as the solution of the equation:
\begin{equation}
\frac{\partial G}{\partial \eta} = 0,
\end{equation}
which according to ($\ref{G}$) reduces to
\begin{equation}
\label{etaLambda}
\eta = g N^{-1} \left<\sigma_{z}\right>_{\tilde{h}_{M}},
\end{equation}
where
\begin{equation}
\tilde{h}_{M} = H_{\text{BCS}} - g\sum_{\alpha} \left( \nu_{\alpha} - \eta\right)\sigma_{z},
\end{equation}
and
\begin{equation}
\left<A\right>_H := \frac{\text{Tr}\left(A\exp\left[-\beta H \right]\right)}{\text{Tr}\exp\left[-\beta H \right]}.
\end{equation}
The solution $\eta(\nu_{1}, \cdots, \nu_{M})$ of equation (\ref{etaLambda}) is unique. The proof can be found in Refs.~\onlinecite{TC1, TC2}. For this choice of $\eta$, the condition for the minimum in (\ref{fLowerBound})
\[
\frac{\partial G}{\partial\nu_{\lambda}} = 0, \quad \lambda=1,\dots,M,
\]
simplifies to
\begin{equation}
\label{nuLambda}
\begin{split}
x^{-1} \nu_{\lambda} = & - \frac{g}{N}\frac{\text{Tr}S_{z\lambda} \exp \left[ -\tfrac{1}{2N}\beta g^{2}S_{z\lambda}^{2} - \beta g S_{z\lambda} \nu_{\lambda}\right]}{\text{Tr} \exp \left[ -\tfrac{1}{2N}\beta g^{2}S_{z\lambda}^{2} - \beta g S_{z\lambda} \nu_{\lambda}\right]} \\& + gN^{-1} \left<\sigma_{z}\right>_{\tilde{h}_{M}}, \quad \lambda = 1,\dots,M.
\end{split}
\end{equation}
The form of these equations is independent of $\lambda$, thus
\begin{equation}
\nu_{1} = \dots = \nu_{M} = \nu.
\end{equation} 
As a consequence, Eqs.~(\ref{nuLambda}) simplify to the form
\begin{equation}
\begin{split}
x^{-1} \nu = & - \frac{g}{N}\frac{\text{Tr}S_{z} \exp \left[ -\tfrac{1}{2N}\beta g^{2}S_{z}^{2} - \beta g S_{z} \nu \right]}{\text{Tr} \exp \left[ -\tfrac{1}{2N}\beta g^{2}S_{z}^{2} - \beta g S_{z} \nu \right]} \\& + gN^{-1} \left<\sigma_{z}\right>_{\tilde{h}},
\end{split}
\end{equation}
where
\begin{equation}
\tilde{h}(\nu, \eta) = H_{\text{BCS}} - g M (\nu - \eta) \sigma_{z}.
\end{equation}

From Eqs.~(\ref{G}) and (\ref{fLowerBound}) one obtains a suitable lower bound to the system's free energy: 
\begin{equation}
\label{fLowerBound2}
|\Lambda| f(H,\beta) \geq - \min_{\nu} \beta^{-1} G(\nu) + \tfrac{1}{2}\beta^{-1} M \ln \left(1 - x^{-1}\right), x > 1,
\end{equation}
where $G(\nu)$ denotes the function defined in (\ref{G}) for $\nu_{1} = \cdots = \nu_{M} = \nu$ and $\eta(\nu)$ satisfies Eq.~(\ref{etaLambda}).

\section{Upper bound to the free energy by Bogolyubov's method}
\label{Sec:UpperBound}
An upper bound on $f(H, \beta)$ can be expressed in terms of the hamiltonian $h^{(M)}(\nu, \eta)$, which is related to $H^{(M)}$ by the following formulae
\begin{equation}
\label{H_M}
H^{(M)} = h^{(M)}(\nu, \eta) + H_{R}^{(M)},
\end{equation}
where
\begin{equation}
\label{hM}
h^{(M)}(\nu, \eta) = \tilde{h} + h_{imp}^{(M)} + \tfrac{1}{2}MN(\nu^{2} - \eta^{2}),
\end{equation}
\begin{equation}
\label{hEl}
\tilde{h} = H_{\text{BCS}} + \kappa \sigma_{z}, \qquad \kappa = - g M (\nu - \eta)
\end{equation}
\begin{equation}
\label{hImp}
h_{imp}^{(M)} = g \nu \sum_{\alpha}S_{z\alpha} + \tfrac{1}{2}N^{-1}g^{2}\sum_{\alpha}S_{z\alpha}^{2},
\end{equation}
\begin{equation}
\label{HR_M}
\begin{split}
H_{R}^{(M)} = & - \tfrac{1}{2}N^{-1} \sum_{\alpha} \Bigl[g\bigl(\sigma_{z} - S_{z\alpha}\bigr) - \nu N\Bigr]^{2} \\& + \tfrac{1}{2} N^{-1} \sum_{\alpha}\bigl(g \sigma_{z} - \eta N\bigr)^{2}.
\end{split}
\end{equation}
\noindent Bogolyubov's inequality
\begin{equation}
\label{Bogolubow}
F(H_{1}+H_{2}) \leq F(H_{1}) + \left<H_{2}\right>_{H_{1}},
\end{equation}
with $H_{1} = h^{(M)}(\nu, \eta)$, yields
\begin{equation}
\label{FF}
\begin{split}
F(H^{(M)}, \beta) & \leq  F(h^{(M)}(\nu, \eta), \beta) \\& + \tfrac{1}{2} N^{-1}\sum_{\alpha} \left<(g{\sigma_{z}} - \eta N)^{2} \right>_{h^{(M)}}.
\end{split}
\end{equation}
The inequality $\text{Tr}(\rho A^{2}) \leq \left(\text{Tr}(\rho A)\right)^{2},$ valid for any bounded self-adjoint operator $A$ and density matrix $\rho$, allows to bound from above the second term in (\ref{FF}) 
\[
\left<(g{\sigma_{z}} - \eta N)^{2} \right>_{h^{(M)}} \leq g^{2}\left<\sigma_{z}\right>_{h^{(M)}}^{2} - 2 g \eta N \left<\sigma_{z}\right>_{h^{(M)}} + \eta^{2} N^{2}.
\]
From Eq. (\ref{etaLambda}) one obtains
\[
\left<(g{\sigma_{z}} - \eta N )^{2} \right>_{h^{(M)}} \leq \eta^{2}N^{2} - 2 \eta^{2}N^{2} + \eta^{2}N^{2} = 0,
\]
which yields the relevant upper bound to the free energy
\begin{equation}
\label{fUpperBound}
F(H^{(M)}, \beta) \leq F(h^{(M)}(\nu, \eta), \beta) = - \beta^{-1} G(\nu).
\end{equation}

\section{Thermodynamic equivalence of $H^{(M)}$~and~$h^{(M)}$}
\label{Sec:BoundsEquality}
By passing to the limit $|\Lambda| \rightarrow \infty$ in Eqs. (\ref{fLowerBound2}), (\ref{fUpperBound}), and subsequently $x \rightarrow 1 + \varepsilon,\ \varepsilon > 0$ in Eq.~(\ref{fLowerBound2}), one finds that the upper and lower bound on $f(H, \beta)$ coalesce almost exactly, if the impurity density is sufficiently small, viz.,
\begin{equation}
\label{limit}
\lim_{|\Lambda| \rightarrow \infty} \text{d-lim }f(H^{(M)}, \beta) \approx \lim_{|\Lambda| \rightarrow \infty} \text{d-lim }f(h^{(M)}(\nu_{\text{m}}, \eta_{\text{m}}), \beta),
\end{equation}
where $\text{d-lim}$ denotes the limit of small $c$ and $\nu_{\text{m}}$, $\eta_{\text{m}}$ are the minimizing solutions of the equations
\begin{equation}
\label{nu}
\nu = gN^{-1} \left<\sigma_{z}\right>_{\tilde{h}} - \frac{g}{N}\frac{\text{Tr}S_{z} \exp \left[ -\tfrac{1}{2N}\beta g^{2}S_{z}^{2} - \beta g S_{z} \nu \right]}{\text{Tr} \exp \left[ -\tfrac{1}{2N}\beta g^{2}S_{z}^{2} - \beta g S_{z} \nu \right]},
\end{equation}
\begin{equation}
\label{eta}
\eta = g N^{-1}\left<\sigma_{z}\right>_{\tilde{h}}.
\end{equation}
On these grounds we shall assume that the thermodynamics of the original system, characterized by $H^{(M)}$ is equivalent, under these restrictions, to that of $h^{(M)}$. 

The two equations (\ref{nu}), (\ref{eta}) reduce to a single one for $\nu$ if $g > 0$. The general form of Eqs.~(\ref{nu}), (\ref{eta}) is
\begin{equation}
\label{Eq:NuGen}
\nu = f_{1}(\nu - \eta) + f_{2}(\nu),
\end{equation}
\begin{equation}
\label{Eq:EtaGen}
\eta = f_{1}(\nu - \eta).
\end{equation}
Let $g > 0$, then $f_{2} > 0$. Furthermore,
\begin{equation}
\label{eta-nu}
\eta = \nu - f_{2}(\nu),
\end{equation}
which yields the equation for $\nu$:
\begin{equation}
\label{nudef}
\nu = f_{1}(f_{2}(\nu)) + f_{2}(\nu),
\end{equation}
where according to Eqs.~(\ref{Eq:NuGen}),~(\ref{Eq:EtaGen}):
\begin{equation}
\label{f1def}
f_{1}(\nu) = (M N\beta)^{-1} \frac{\partial}{\partial \nu} \ln \text{Tr} \exp[ - \beta \tilde{h}(\nu,0)],
\end{equation}
\begin{equation}
\label{f2def}
f_{2}(\nu) = (N\beta)^{-1} \frac{\partial}{\partial \nu} \ln \text{Tr} \exp[ - \beta h^{(1)}_{imp}].
\end{equation}

\section{Mean-field description of $h$}
\label{Sec:FreeEnergyMinimization}
The form of the Hamiltonian $\tilde{h}$ (\ref{hEl}) is analogous to 
\begin{equation}
H_{\text{BCS}}({\mathcal H}) = H_{0} + V_{\text{BCS}} - \mu_{\text{B}} {\mathcal H} \sigma_{z}, 
\end{equation}
describing a system of electrons with attractive BCS interaction in the presence of an external magnetic field $\mathcal H$ ($\mu_{\text{B}}$ denotes the Bohr magneton). 
The explicit form of the system's free energy  $f(h^{(M)}(\nu, \eta), \beta)$ can be therefore derived by exploiting the Bogolubov-Valatin transformation\cite{Bogolyubov, Valatin} and the method developed in Ref.~\onlinecite{Rickayzen} for $H_{\text{BCS}}({\mathcal H})$. The Bogolubov-Valatin transformation,
\begin{eqnarray}
\label{BVtransormation}
\begin{array}{c}
\alpha^{}_{\mathbf{k}+} = u^{}_{\mathbf{k}} a^{}_{\mathbf{k}+} - v^{}_{\mathbf{k}} a^{\dagger}_{-\mathbf{k}-}, \\
\alpha^{}_{\mathbf{k}-} = u^{}_{\mathbf{k}} a^{}_{\mathbf{k}-} + v^{}_{\mathbf{k}} a^{\dagger}_{-\mathbf{k}+},
\end{array}
\end{eqnarray}
yields
\begin{equation}
\label{sigmaAlpha}
\sigma_{z} = \sum_{\mathbf{k}} (\alpha_{\mathbf{k}+}^{\dagger}\alpha_{\mathbf{k}+}^{} -  \alpha_{-\mathbf{k}-}^{\dagger}\alpha_{-\mathbf{k}-}^{}).
\end{equation}
One expects the energies of the states $\mathbf{k}+,$ $-\mathbf{k}-$ to be different, since the products $\alpha_{\mathbf{k}+}^{\dagger}\alpha_{\mathbf{k}+}^{}$, $\alpha_{-\mathbf{k}-}^{\dagger}\alpha_{-\mathbf{k}-}^{}$ appear with the opposite signs in  Eq.~(\ref{sigmaAlpha}). Thus, the trial equilibrium density matrix, approximating $\tilde{Z}^{-1} \exp[-\beta \tilde{h}]$, has the form (see Ref.~\onlinecite{Rickayzen})
\begin{equation}
\label{rho0}
\rho_{0} = \frac{\exp[-\beta \tilde{H_{0}]}}{\text{Tr}\exp[-\beta \tilde{H_{0}}]},
\end{equation}
is characterized by the following Hamiltonian
\begin{equation}
\label{h0}
\tilde{H_{0}} = \sum_{\mathbf{k}} (E_{\mathbf{k}1} \alpha_{\mathbf{k}+}^{\dagger}\alpha_{\mathbf{k}+}^{} + E_{\mathbf{k}2}  \alpha_{-\mathbf{k}-}^{\dagger}\alpha_{-\mathbf{k}-}^{}) + E_{0},
\end{equation}
where $E_{0}$ denotes the ground state energy.
According to (\ref{hM}) and (\ref{limit}) the system's free energy can be decomposed into three summands 
\begin{equation}
\label{FreeEnergy}
F = F_{el} + F_{imp} + \tfrac{1}{2}MN \left(\nu^{2} - \eta^{2}\right),
\end{equation}
where
\begin{equation}
F_{imp} = - \beta^{-1} \ln \text{Tr} \exp[-\beta h_{imp}],
\end{equation}
is the free energy of impurities and 
\begin{equation}
F_{el} =  U_{el} - TS_{el},
\end{equation}
denotes free energy of electrons, with 
\begin{equation}
U_{el} = \text{Tr} (\tilde{h} \rho_{0}) 
\end{equation}
and
\begin{equation}
S_{el} = -(T \beta)^{-1} \text{Tr} (\rho_{0} \ln \rho_{0}).
\end{equation}
For $\rho_{0}$ defined by Eq.~(\ref{rho0}),
\begin{equation}
\label{EntropyBV}
\begin{split}
S_{el} = & -(\beta T)^{-1} \sum_{\mathbf{k}} \Bigl[f_{\mathbf{k}1} \ln f_{\mathbf{k}1} + (1 - f_{\mathbf{k}1}) \ln (1 - f_{\mathbf{k}1}) \\& + f_{\mathbf{k}2} \ln f_{\mathbf{k}2} + (1 - f_{\mathbf{k}2}) \ln (1 - f_{\mathbf{k}2})\Bigr],
\end{split}
\end{equation}
where
\begin{equation}
\label{fk_def}
f_{\mathbf{k}1} = \frac{\exp(-\beta E_{\mathbf{k}1})}{1+\exp(-\beta E_{\mathbf{k}1})}, \qquad f_{\mathbf{k}2} = \frac{\exp(-\beta E_{\mathbf{k}2})}{1+\exp(-\beta E_{\mathbf{k}2})}.
\end{equation}
The projectors on the ground and excited BCS states
\begin{displaymath}
\begin{array}{c}
P_{0\mathbf{k}} = \left|BCS\right>_{\mathbf{k\ k}}\left<BCS\right|, \\
P_{\mathbf{k}+} = \alpha^{\dagger}_{\mathbf{k}+} \left|BCS\right>_{\mathbf{k\ k}}\left<BCS\right| \alpha^{}_{\mathbf{k}+}, \\ P_{-\mathbf{k}-} = \alpha^{\dagger}_{-\mathbf{k}-} \left|BCS\right>_{\mathbf{k\ k}} \left<BCS\right| \alpha^{}_{-\mathbf{k}-}, \\
P_{\mathbf{k}+,-\mathbf{k}-} = \alpha^{\dagger}_{\mathbf{k}+} \alpha^{\dagger}_{-\mathbf{k}-} \left|BCS\right>_{\mathbf{k\ k}} \left<BCS\right| \alpha^{}_{-\mathbf{k}-} \alpha^{}_{\mathbf{k}+},
\end{array}
\end{displaymath}
allow to rewrite Eq.~(\ref{rho0}) in the following form
\begin{equation}
\label{rho0expl}
\begin{split}
\rho_{0} = & \prod_{\mathbf{k}} \Bigl( (1-f_{\mathbf{k1}})(1-f_{\mathbf{k}2}) P_{0\mathbf{k}} + f_{\mathbf{k}1} (1- f_{\mathbf{k}2})P_{\mathbf{k}+} \\& + f_{\mathbf{k}2} (1- f_{\mathbf{k}1}) P_{-\mathbf{k}-} + f_{\mathbf{k}1}f_{\mathbf{k}2}P_{\mathbf{k}+,-\mathbf{k}-} \Bigr).
\end{split}
\end{equation}
Thus, one obtains
\begin{equation}
\label{InternalEnergyBV}
\begin{split}
U_{el} &= \sum_{\mathbf{k}} \Bigl[f_{\mathbf{k}1}(\xi_{\mathbf{k}} + \kappa) + f_{\mathbf{k}2}(\xi_{\mathbf{k}} - \kappa) + 2\xi_{\mathbf{k}} v_{\mathbf{k}}^{2} (1 - f_{\mathbf{k}1} \\ & - f_{\mathbf{k}2} )\Bigr] - |\Lambda|^{-1}\sum_{\mathbf{k\,k^{\prime}}} G_{\mathbf{k\,k^{\prime}}} u_{\mathbf{k}} v_{\mathbf{k}} u_{\mathbf{k^{\prime}}} v_{\mathbf{k^{\prime}}} \bigl(1 - f_{\mathbf{k}1} - f_{\mathbf{k}2} \bigr) \\ & \times \bigl(1 - f_{\mathbf{k^{\prime}}1} - f_{\mathbf{k^{\prime}}2} \bigr),
\end{split}
\end{equation}
which yields
\begin{equation}
\begin{split}
\label{FreeEnergyBV}
F_{el} & = U_{el} + \beta^{-1} \sum_{\mathbf{k}} \Bigl[f_{\mathbf{k}1} \ln f_{\mathbf{k}1} + (1 - f_{\mathbf{k}1}) \ln (1 - f_{\mathbf{k}1}) \\& + f_{\mathbf{k}2} \ln f_{\mathbf{k}2} + (1 - f_{\mathbf{k}2}) \ln (1 - f_{\mathbf{k}2})\Bigr].
\end{split}
\end{equation}
The parameters $u_{\mathbf{k}}$, $v_{\mathbf{k}}$, $E_{\mathbf{k}1}$ and $E_{\mathbf{k}2}$, are found by minimizing the free energy. 
The thermodynamic perturbation method of Bogolyubov et al.\cite{BogolubowZubarevTsernikov} shows that for such choice of these parameters, the following relation holds up to negligible terms
\begin{equation}
\label{Limith0}
\lim_{|\Lambda| \rightarrow \infty} f (\tilde{h}, \beta) = \lim_{|\Lambda| \rightarrow \infty} \min_{\{\Delta_{\mathbf{p}}\}} f(\tilde{h}_{0}, \beta),
\end{equation} 
where $\tilde{h}_{0} = \tilde{H_{0}} + \kappa \sigma_{z}$ and 
\begin{equation}
\label{DeltaDef}
\Delta_{\mathbf{p}} = |\Lambda|^{-1} \sum_{\mathbf{k}} G_\mathbf{p\,k} u_{\mathbf{k}} v_{\mathbf{k}} \bigl(1 - f_{\mathbf{k}1} - f_{\mathbf{k}2} \bigr).
\end{equation}
Minimization of Eq.~(\ref{FreeEnergyBV}) with respect to $v_{\mathbf{p}}$ yields
\begin{equation}
\label{Minimization}
\begin{split}
\frac{\partial F_{el}}{\partial v_{\mathbf{p}}} & = 4 \xi_{\mathbf{p}} v_{\mathbf{p}} \bigl(1 - f_{\mathbf{p}1} - f_{\mathbf{p}2} \bigr) - 2 |\Lambda|^{-1}\sum_{\mathbf{k}} G_\mathbf{p\,k} u_{\mathbf{k}} v_{\mathbf{k}} \\& \bigl(u_{\mathbf{p}} - \frac{v_{\mathbf{p}}^2}{u_{\mathbf{p}}}\bigr) \bigl(1 - f_{\mathbf{k}1} - f_{\mathbf{k}2} \bigr) \bigl(1 - f_{\mathbf{p}1} - f_{\mathbf{p}2} \bigr) = 0.
\end{split}
\end{equation}
Then, from the normalization condition $u_{\mathbf{k}}^{2} + v_{\mathbf{k}}^{2} = 1$ and the demand that the density matrix $\rho_{0}$ should represent the Fermi-Dirac distribution of free fermions for $\Delta = 0$, one obtains
\begin{equation}
\label{amplitudes}
u_{\mathbf{k}}^{2} = \frac{1}{2} \Bigl(1 + \frac{\xi_{\mathbf{k}}}{E_{\mathbf{k}}}\Bigr), \quad v_{\mathbf{k}}^{2} = \frac{1}{2} \Bigl(1 - \frac{\xi_{\mathbf{k}}}{E_{\mathbf{k}}}\Bigr), \quad E_{\mathbf{k}} = \sqrt{\xi_{\mathbf{k}}^{2} + \Delta_{\mathbf{k}}^{2}}.
\end{equation}
The equation
\[
\begin{split}
\frac{\partial F_{el}}{\partial f_{\mathbf{k}1}} & = \xi_{\mathbf{k}} + \kappa - 2\xi_{\mathbf{k}} v_{\mathbf{k}}^{2} + 2|\Lambda|^{-1} \sum_{\mathbf{k^{\prime}}} G_{\mathbf{k\,k}^{\prime}}u_{\mathbf{k}} v_{\mathbf{k}} u_{\mathbf{k^{\prime}}} v_{\mathbf{k^{\prime}}} \\ & \times \bigl(1 - f_{\mathbf{k^{\prime}}1} - f_{\mathbf{k^{\prime}}2} \bigr) + \beta^{-1} \bigl(\ln f_{\mathbf{k}1} - \ln(1 - f_{\mathbf{k}1}) \bigr) = 0,
\end{split}
\]
combined with Eqs.~(\ref{fk_def}), (\ref{DeltaDef}), yields
\begin{equation}
\label{Ek1}
E_{\mathbf{k}1} = \xi_{\mathbf{k}} (u_{\mathbf{k}}^{2} - v_{\mathbf{k}}^{2}) + \kappa + 2 \Delta_{\mathbf{k}} u_{\mathbf{k}} v_{\mathbf{k}} = E_{\mathbf{k}} + \kappa.
\end{equation}
Analogously, for $E_{\mathbf{k}2}$ one obtains
\begin{equation}
\label{Ek2}
E_{\mathbf{k}2} = \xi_{\mathbf{k}} (u_{\mathbf{k}}^{2} - v_{\mathbf{k}}^{2}) - \kappa + 2 \Delta_{\mathbf{k}} u_{\mathbf{k}} v_{\mathbf{k}} = E_{\mathbf{k}} - \kappa.
\end{equation}
Using equations (\ref{eta-nu}), (\ref{fk_def}), (\ref{amplitudes}) we get the equation for the gap parameter
\begin{equation}
\label{Delta_k}
\Delta_{\mathbf{k}} = \tfrac{1}{2} |\Lambda|^{-1} \sum_{\mathbf{k}^{\prime}} G_{\mathbf{k\,k}^{\prime}} \Delta_{\mathbf{k}^{\prime}} E_{\mathbf{k}^{\prime}}^{-1} f_{3}(\beta, E_{\mathbf{k}^{\prime}}, \xi_{\mathbf{k}^{\prime}}, f_{2}),
\end{equation}
where
\begin{equation}
\label{f3}
f_{3}(\beta, E_{\mathbf{k}^{\prime}}, \xi_{\mathbf{k}^{\prime}}, f_{2}) = \frac{\sinh(\beta E_{\mathbf{k}^{\prime}})}{\cosh(\beta E_{\mathbf{k}^{\prime}}) + \cosh(g \beta M f_{2}(\nu))}.
\end{equation}
Eq.~(\ref{Delta_k}) resembles the gap equation in BCS theory. The convexity properties of $f_{3}(\beta, E_{\mathbf{k}^{\prime}}, \xi_{\mathbf{k}^{\prime}}, f_{2})$ differ in general from those of $f_{\text{BCS}} = \tanh(\tfrac{1}{2}\beta E_{\mathbf{k}})$. However, in the limit of extremely weak magnetic coupling, $g \rightarrow 0$, $f_{3}$ reduces to $f_{\text{BCS}}$, viz., 
\begin{equation}
\begin{split}
f_{3}\bigr|_{g=0} & = \frac{\sinh(\beta E_{\mathbf{k}^{\prime}})}{\cosh(\beta E_{\mathbf{k}^{\prime}}) + 1} \\& = \frac{2\sinh(\tfrac{1}{2}\beta E_{\mathbf{k}^{\prime}})\cosh(\tfrac{1}{2}\beta E_{\mathbf{k}^{\prime}})}{2\cosh^{2}(\tfrac{1}{2}\beta E_{\mathbf{k}^{\prime}})} \\& = \tanh(\tfrac{1}{2}\beta E_{\mathbf{k}^{\prime}}) \\& = f_{\text{BCS}}.
\end{split}
\end{equation}
Equalities (\ref{InternalEnergyBV}), (\ref{FreeEnergyBV}), (\ref{amplitudes}), (\ref{Delta_k}) and (\ref{f3}) now lead to the following expression for the free energy
\begin{equation}
\label{ElFreeEnergy}
\begin{split}
F_{el} & = \sum_{\mathbf{k}} \Bigl[\xi_{\mathbf{k}} + \tfrac{1}{2} \Delta_{\mathbf{k}}^{2} E_{\mathbf{k}}^{-1} f_{3}(\beta, E_{\mathbf{k}}, \xi_{\mathbf{k}}, f_{2}) \\& - \beta^{-1} \ln \bigl[2 \cosh(\beta E_{\mathbf{k}}) + 2 \cosh(\beta \kappa)\bigr]\Bigr].
\end{split}
\end{equation}
The definition (\ref{f1def}), where $\eta = 0$, together with Eq.~(\ref{ElFreeEnergy}), yields the explicit form of the function $f_{1}(\nu)$:
\begin{equation}
\label{f1}
f_{1}(\nu) = \frac{cg}{M} \frac{\sinh(g \beta M \nu)}{\cosh(g \beta M \nu) + \cosh(\beta E_{\mathbf{k}})}.
\end{equation}

The free energy of impurities and the function $f_{2}$ depend on the value of their spin. In the present work we examine the critical temperature $T_{\text{c}}$ of alloys containing Ce and Gd magnetic impurities. According to Matthias et al. \cite{Matthias1} the spin of Ce ion is $1/2$ and that of Gd is $7/2$. For spin $1/2$ impurities one obtains,
\begin{equation}
\label{Fimp_1_2}
\begin{split}
F_{imp}^{(\tfrac{1}{2})} &= -\beta^{-1} \ln \text{Tr} \exp \bigl[-\beta h_{imp}^{(M)}\bigr] \\& = -\beta^{-1} \sum_{\alpha = 1}^{M} \ln \text{Tr} \exp \bigl[-\beta h_{imp}^{(1)}\bigr] \\&= - M \beta^{-1} \ln \bigl[2 \cosh(\beta g \nu)\bigr] + \tfrac{1}{2} c g^{2},
\end{split}
\end{equation}
and
\begin{equation}
\label{f2_1_2}
f_{2}^{(\tfrac{1}{2})}(\nu) = \frac{c g}{M} \tanh(\beta g \nu).
\end{equation}
\begin{widetext}
Accordingly, for spin $7/2$ impurities
\begin{equation}
\label{Fimp_7_2}
\begin{split}
F_{imp}^{(\tfrac{7}{2})} = & - M \beta^{-1} \ln 2 \Bigl[\exp[-24 \tfrac{g^{2} \beta}{N}] \cosh(7 \beta g \nu) + \exp[-12 \tfrac{g^{2} \beta}{N}] \cosh(5 \beta g \nu) \\&+ \exp[-4\tfrac{g^{2} \beta}{N}] \cosh(3 \beta g \nu) + \cosh( \beta g \nu) \Bigr] + \tfrac{1}{2}cg^{2},
\end{split}
\end{equation}
with
\begin{equation}
\label{f2_7_2}
\begin{split}
f_{2}^{(\tfrac{7}{2})}(\nu) & = \frac{c g}{M R} 7 \exp[-24 \frac{g^{2} \beta}{N}] \sinh(7 \beta g \nu) + 5 \exp[-12 \frac{g^{2} \beta}{N}] \sinh(5 \beta g \nu) + 3\exp[-4 \frac{g^{2} \beta}{N}] \sinh(3 \beta g \nu) + \sinh(\beta g \nu),
\end{split}
\end{equation}
where
\[
\begin{split}
R &= \exp[-24 \tfrac{g^{2} \beta}{N}] \cosh(7 \beta g \nu) + \exp[-12 \tfrac{g^{2} \beta}{N}] \cosh(5 \beta g \nu) + \exp[-4\tfrac{g^{2} \beta}{N}] \cosh(3 \beta g \nu)  + \cosh( \beta g \nu).
\end{split}
\]
\end{widetext}
\section{The chemical potential}
\label{Sec:ChemicalPotential}
Passing from summation in Eq.~(\ref{Delta_k}) over momentum $\mathbf{k}$ to integration over the single electron energies $\xi$, one obtains for a sufficiently thin conduction band $S$:
\begin{equation}
\label{Delta}
\Delta = \frac{1}{2} G_{0} \rho \int_{-\delta}^{\delta} \frac{\Delta}{E}f_{3}\bigl(\beta, E, \xi, f^{(s)}_{2}(\nu)\bigr)\text{d}\xi \qquad s = 1/2,\ 7/2,
\end{equation}
where $\rho$ denotes the density of states in ${\mathcal P}$.
This equation together, with the one for $\nu$,
\begin{equation}
\label{nuEq}
\begin{split}
\nu & = f_{1}\bigl(f^{(s)}_{2}(\nu)\bigr) + f^{(s)}_{2}(\nu) \\& = \frac{cg}{M} \frac{\sinh\bigl(\beta g M f^{(s)}_{2}(\nu)\bigr)}{\cosh\bigl(\beta g M f^{(s)}_{2}(\nu)\bigr) + \cosh(\beta E_{\mathbf{k}})} + f^{(s)}_{2}(\nu),
\end{split}
\end{equation}
constitutes the set of equations for $\Delta$ and $\nu$. The properties of a superconductor with magnetic impurities can be determined by solving this set of equations, which is supplemented by the following condition for the chemical potential $\mu$:
\begin{equation}
\label{ChemicalPotential}
\sum_{\mathbf{k}\sigma}\text{Tr}\bigl(n_{\mathbf{k}\sigma} \rho_{0} \bigr) = n,
\end{equation}
$n$ denoting the average number of fermions in the system. According to Eqs.~(\ref{BVtransormation}), (\ref{fk_def}), (\ref{rho0expl}) and (\ref{f3}), this condition, takes the form:
\begin{equation}
\label{mu}
\sum_{\mathbf{k}} \Bigl[1 - \frac{\xi_{\mathbf{k}}}{E_{\mathbf{k}}} f_{3}\bigl(\beta, E_{\mathbf{k}}, \xi_{\mathbf{k}}, f^{(s)}_{2}\bigr) \Bigr] = n.
\end{equation}
Eq.~(\ref{mu}) is analogous to the BCS equation for $\mu$ and the properties of $f_{3}$ are similar to $f_{\text{BCS}} = \tanh(\beta E_{\mathbf{k}}/2)$, e.g. both are odd functions in $\xi_{\mathbf{k}}$. The solution of Eq.~(\ref{mu}) is therefore exactly the same as in BCS theory, viz., $\mu = \varepsilon_{\text{F}}$. Numerical calculations in subsequent sections are thus performed under the following assumptions:
\begin{equation}
\mu = \varepsilon_{\text{F}}, \qquad \frac{\partial \mu}{\partial T} = 0, \qquad \rho = \rho_{\text{F}},
\end{equation}
where $\rho_{F}$ denotes the density of states at the Fermi level,
\[
\rho_{F} = \frac{m p_{F}}{2 \pi^{2} \hbar^{2}}.
\]
We have also imposed the condition
\begin{equation}
\label{ef_ineq}
g M f_{2}(\nu) < E_{\mathbf{k}}\text{ at $T = 0\,\text{K}$,}
\end{equation}
which expresses weak coupling between conduction electrons and impurities. Eq.~(\ref{Delta}) then takes the limiting form
\begin{equation}
\label{Delta0}
G_{0} \rho_{F} \, \text{arcsinh} \left(\frac{\delta}{\Delta(0)}\right) = 1, \text{ as $T\rightarrow 0$,}
\end{equation}
whereas Eq.~(\ref{nuEq}) is satisfied by $\nu(0) = \frac{c g}{M}$ in this limit. Thus at sufficiently low temperatures $T$, close to $0\,\text{K}$, $\Delta(T)$ is the solution of Eq.~(\ref{Delta0}). This is substituted into Eq.~(\ref{nuEq}), which is then solved for one-fermion energies $\xi \in {\mathcal P}$ by exploiting the Newton-Raphson method. The resulting values of $\nu(\xi)$ are used to obtain $\Delta(T + \Delta T)$ from (\ref{Delta}) by deploying Newton-Cotes quadrature as long as the result is self-consistent. The resulting value of $\Delta(T + \Delta T)$ is used to compute $\nu(\xi)$ at temperature $T + \Delta T$. This procedure is continued until $T$ reaches the specified value. To ensure numerical stability the step $\Delta T$ should be sufficiently small, e.g. $\Delta T = 10^{-3}\,\text{K}$. 

In the opposite case, viz., $g M f_{2}(\nu) > E_{\mathbf{k}}$ for $T = 0\,\text{K}$ one has
\[
\Delta(0) = 0 \qquad \nu(0) = \frac{2 c g}{M}.
\]
In such anomalous circumstances, superconductivity is completely suppressed at $T = 0\,\text{K}$ and the system is described only by the impurity parameter $\nu$. The intermediate case, when $g M f_{2}(\nu) = E_{\mathbf{k}}$ has not been examined numerically.

\section{Phase Diagrams}
\label{Sec:PhaseDiagrams}
The system's state is characterized, according to Eqs.~(\ref{limit}) and (\ref{Limith0}), by the solution of Eqs.~(\ref{Delta}) and (\ref{nuEq}), which minimizes $F(h^{(M)}, \beta)$. It will be denoted by $\{\Delta_{\text{m}}, \nu_{\text{m}}\}$.

Passing from summation in Eq.~(\ref{ElFreeEnergy}) over $\mathbf{k}$ to integration over the single-fermion energies $\xi$ and exploiting Eq.~(\ref{FreeEnergy}), one obtains the following expression for the free energy
\begin{widetext}
\begin{equation}
\label{F_1_2}
\begin{split}
F^{(s)} = & \min_{\{\Delta,\,\nu\}} \Bigl\{ \rho_{F} |\Lambda| \int_{-\delta}^{\delta} \Bigl[ \tfrac{1}{2} \Delta^{2} E^{-1} f_{3}\bigl(\beta, E, \xi, f_{2}^{(s)}\bigr) - \beta^{-1} \ln \bigl[2 \cosh(\beta E) + 2 \cosh\bigl(g \beta M f_{2}^{(s)}\bigr)\bigr]\Bigr]\text{d}\xi \\ &+ M^{2}c^{-1} \bigl(\nu f_{2}^{(s)} - \tfrac{1}{2} \bigl(f_{2}^{(s)}\bigr)^{2}\bigr) + F_{imp}^{(s)} + E_{0}(\Delta = 0) + \rho_{F} \delta^{2}\Bigr\}, \qquad s = 1/2,\ 7/2,
\end{split}
\end{equation}
\end{widetext}
where $F_{imp}^{(s)}$ are given by the Eqs.~(\ref{Fimp_1_2}) and (\ref{Fimp_7_2}), whereas $E_{0}(\Delta = 0)$ denotes the ground state energy of free fermions. Two last terms are the contribution
to the free energy density from one-fermion states, lying
outside $S$.

Equations (\ref{Delta}) and (\ref{nuEq}) clearly possess the solution $\Delta = \nu = 0$ for all values of $\beta \geq 0$. At sufficiently large values of $\beta$ one finds also other solutions, viz., $\{ \Delta \neq 0, \nu = 0 \}$, $\{ \Delta = 0, \nu \neq 0 \}$, $\{ \Delta \neq 0, \nu \neq 0 \}$. Accordingly, we distinguish the following phases:
\begin{itemize}
\item[$-$]{paramagnetic phase $P$ with $\{\Delta_{\text{m}} = 0, \nu_{\text{m}} = 0 \}$,}
\item[$-$]{unperturbed superconducting state $SC$ with $\{\Delta_{\text{m}} \neq 0, \nu_{\text{m}} = 0 \}$,}
\item[$-$]{ferromagnetic phase $F$ without bound Cooper pairs and $\{\Delta_{\text{m}} = 0, \nu_{\text{m}} \neq 0 \}$, in which impurity spins tend to align opposite to those of conduction fermions (cf. Eqs.~(\ref{hEl}) and (\ref{hImp})).}
\item[$-$]{intermediate phase $D$ in which superconductivity coexists with ferromagnetism and $\{\Delta_{\text{m}} \neq 0, \nu_{\text{m}} \neq 0 \}$.}
\end{itemize}

We define the following temperatures corresponding to the respective phase transitions 
\begin{itemize}
\item[$-$]{$T_{\text{c}}$, 2nd order transition $SC$ $\rightarrow$ $P$.}
\item[$-$]{$T_{PF}$, Curie temperature of 2nd order transition $F$ $\rightarrow$ $P$.}
\item[$-$]{$T_{SCD}$, 1st order transition $D$ $\rightarrow$ $SC$.}
\item[$-$]{$T_{FD}$, 1st order transition $D$ $\rightarrow$ $F$.}
\item[$-$]{$T_{SCF}$, 1st order transition $SC$ $\rightarrow$ $F$.}
\end{itemize}

The set of Eqs.~(\ref{Delta}), (\ref{nuEq}) has been solved numerically for different values of $g$, $M$, $\delta$ and $G_{0}\rho_{\text{F}}$. The parameters were adjusted to fit the experimental specific-heat curves for (La$_{1-x}$Ce$_{x}$)Al$_{2}$ and LaGd and the critical field curve in case of ThGd\cite{DB10p}. These values are used to compute the phase diagrams of these alloys on the grounds of Eq.~(\ref{F_1_2}).

\begin{figure*}
\begin{center}
\begin{tabular}{@{}c@{ }c@{ }c@{ }c@{}@{ }@{ }c@{ }c@{ }c@{ }c@{ }}
\multicolumn{1}{l}{\footnotesize \bf{a)}} & \multicolumn{1}{l}{\footnotesize \bf{b)}} \\[-0.9cm]
    \includegraphics[width=7.5cm]{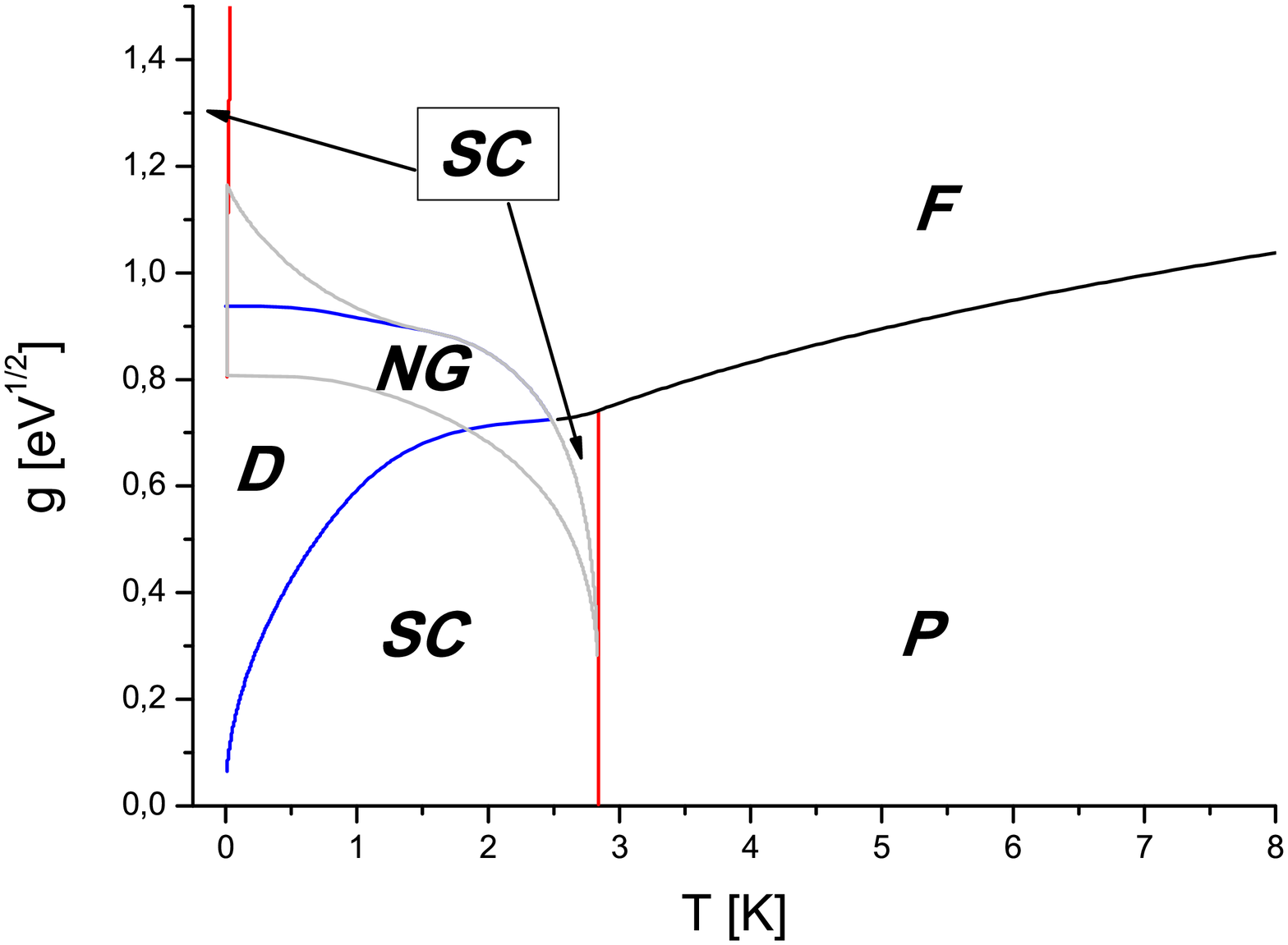} &
    \includegraphics[width=7.5cm]{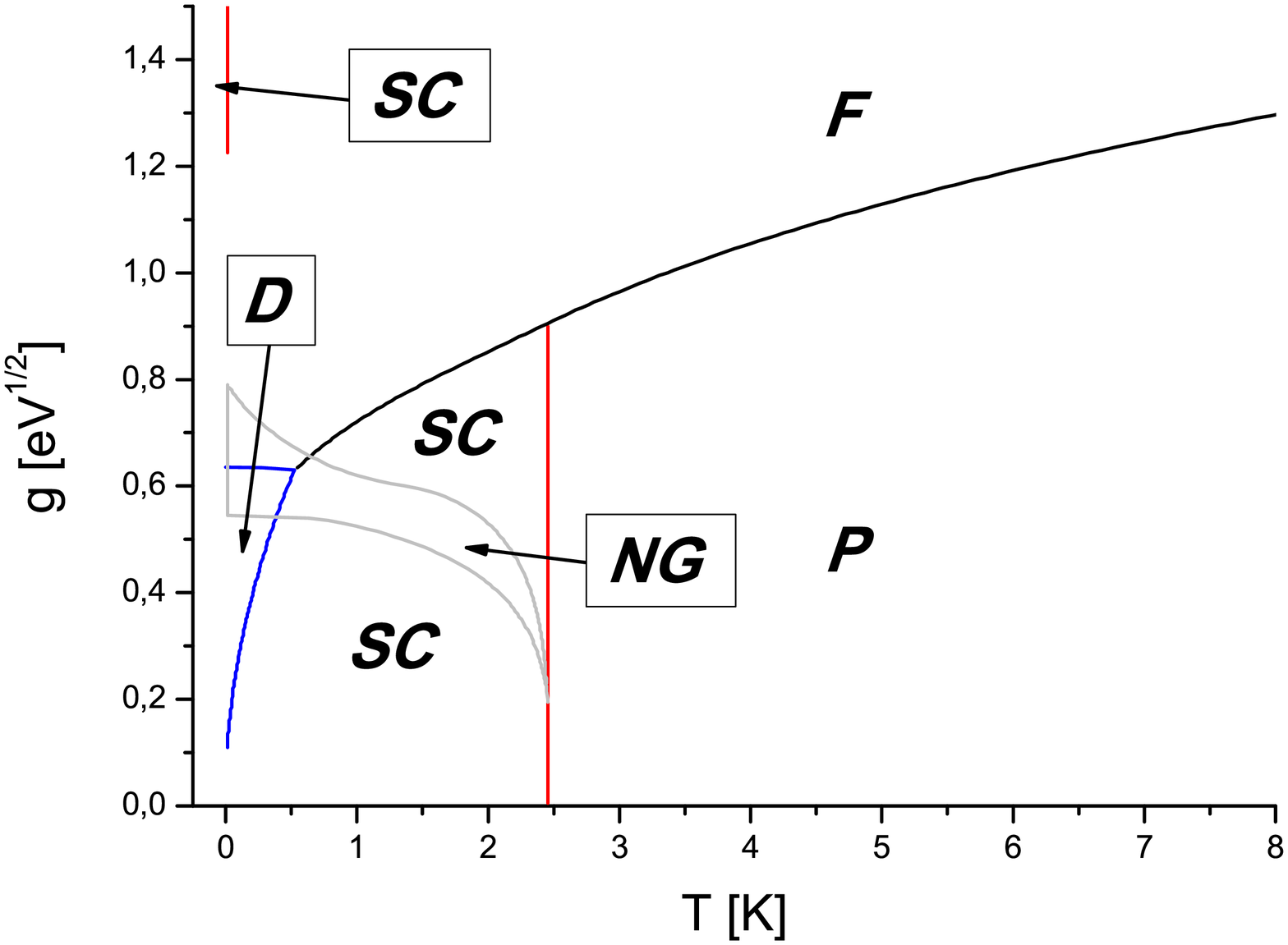}\\
\multicolumn{1}{l}{\footnotesize \bf{c)}} & \multicolumn{1}{l}{\footnotesize \bf{d)}} \\[-0.9cm]
    \includegraphics[width=7.5cm]{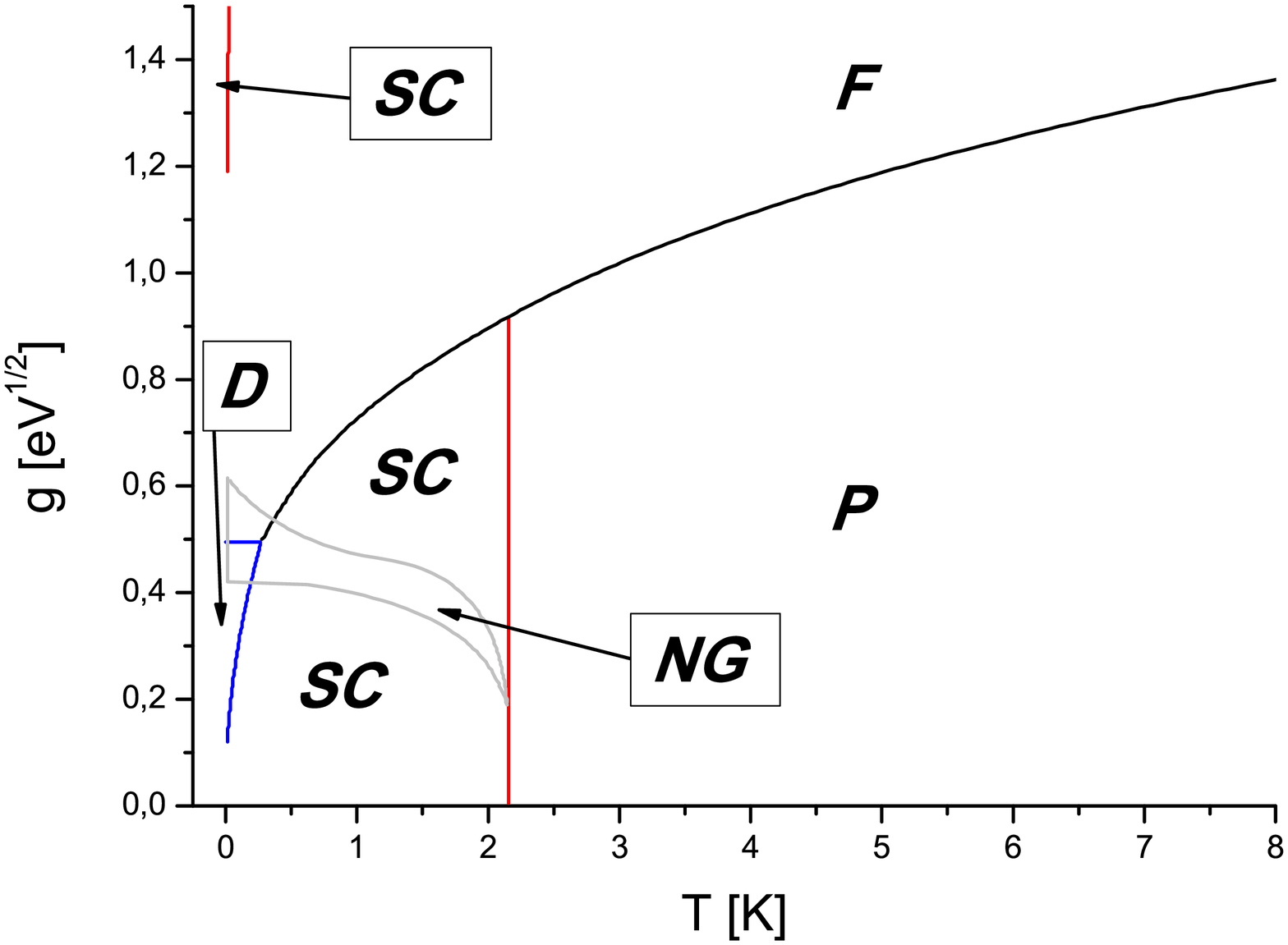}&
    \includegraphics[width=7.5cm]{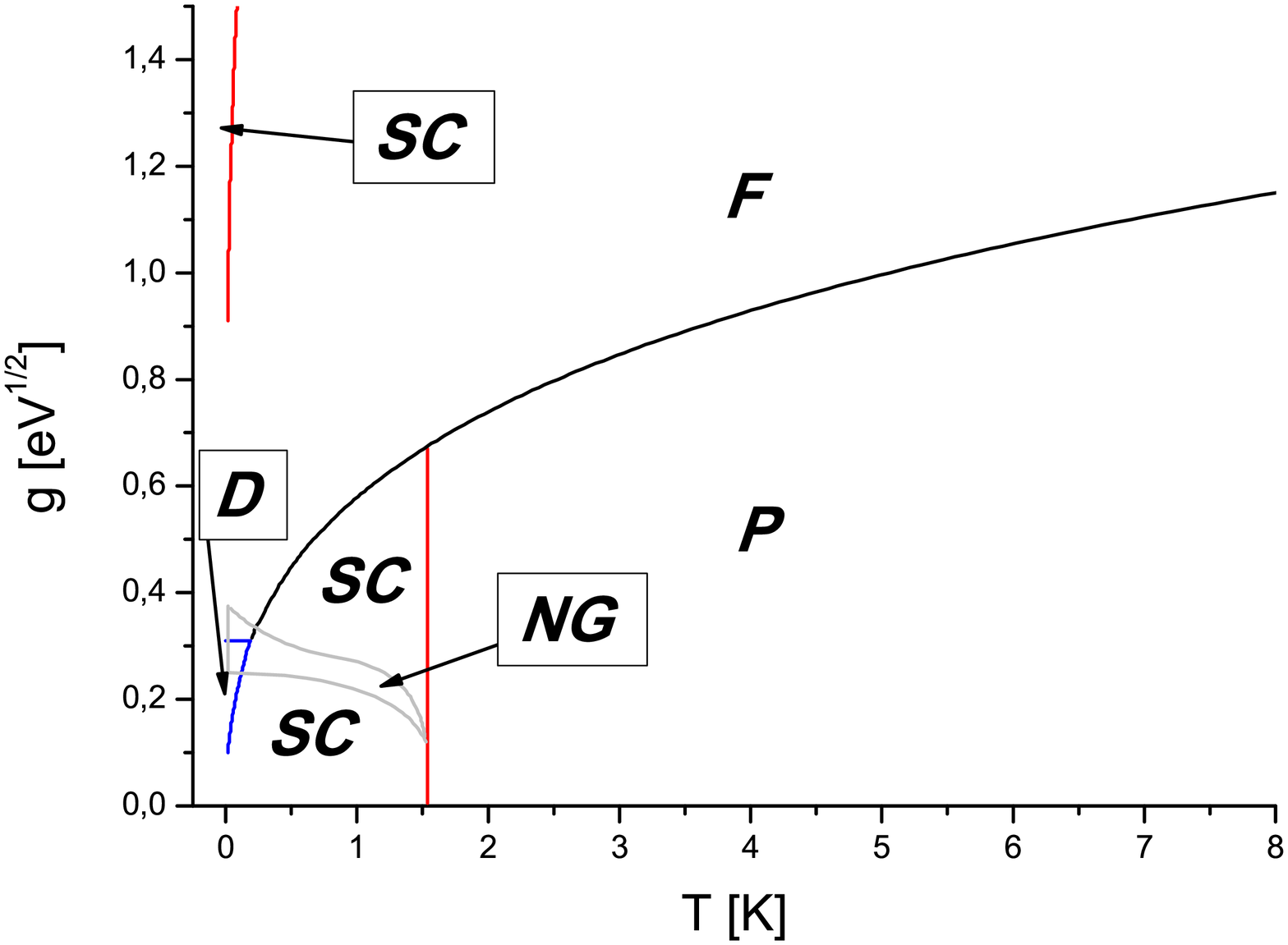}\\
\end{tabular}
\caption{\label{Fig1}(Color online) Phase diagrams of (La$_{1-x}$Ce$_{x})$Al$_{2}$ for the values of $M$, $g$, $\delta$ and $G_{0}\rho_{F}$ given in Table~\ref{table1} and under varying $x$: (a) $x = 0.0010$, (b) $x = 0.0019$, (c) $x = 0.0028$ and (d) $x = 0.0057$. The following convention was used in order to distinguish the regions corresponding to different phases: $SC$ (red), $D$ (blue), $F$ (black) and the gapless phase $NG$ (grey), where the smallest excitation energies from ground state disappear.}
\end{center}
\end{figure*}

The phase diagrams of (La$_{1-x}$Ce$_{x}$)Al$_{2}$, for which the free energy is given by Eq.~(\ref{F_1_2}) with $s=1/2$ and impurity concentration $c=x/(3-x)$ are depicted in Fig.~\ref{Fig1}. The values of $M$, $g$, $\delta$ and~$G_{0}\rho_{F}$ are given in Table~\ref{table1}. These diagrams show the decline of $D$ and $SC$ phase with increasing impurity concentration. The critical temperature $T_{\text{c}}$ is a decreasing function of impurity concentration. It follows, therefore, that the destructive effect of impurities increases with $c$ and suppresses the superconductivity if $g$ or $c$ reaches its critical value. This complies with experimental data, in which superconductivity is expunged for Ce content larger than $0.67\%$\cite{MapleConvAndUnconvSC, WF75, SB75}. 

The supplementary phase $NG$ with gapless superconductivity and $\Delta_{\text{m}} > 0$ is present, depending on the value of $g$, as a subregion of $SC$, $D$ or $F$ phase. The appearance of $NG$ subregion is due to negative term $-g M f_{2}(\nu)$ present in $E_{\mathbf{k}1}$ (\ref{Ek1}).

For sufficiently small magnetic coupling constant $g$, the $NG$ phase for (La$_{1-x}$Ce$_{x})$Al$_{2}$ lies between $D$ and $F$ phases. Thus, the pair-breaking mechanism, which increases with $g$, induces gapless superconductivity and then superconductivity is suppressed for sufficiently large $g$. The system undergoes a phase transition to a ferromagnetic $F$ state (cf. Fig.~\ref{Fig1}a). 

\begin{table*}
\tabcolsep12pt
\begin{center}
\begin{tabular}{*{8}{|c}|}
\hline
Alloy & $x$ & c [\%] & $M$ & $g\ [\sqrt{\text{eV}}]$ & $\delta\ [\text{eV}]$ & $G_{0} \rho_{F}$\\
\hline\hline
(LaCe)Al$_{2}$ & $0.0010$ & $-$ & $1$ & $0.10$ & $0.01$ & $0.2610$ \\
(LaCe)Al$_{2}$ & $0.0019$ & $-$ & $4$ & $0.189$ & $0.01$ & $0.2515$ \\
(LaCe)Al$_{2}$ & $0.0028$ & $-$ & $7$ & $0.19$ & $0.01$ & $0.2435$ \\
(LaCe)Al$_{2}$ & $0.0057$ & $-$ & $8$ & $0.23$ & $0.01$ & $0.2250$\\
\hline\hline
LaGd & $-$ & $0.20$ & $5$ & $0.16$ & $0.01$ & $0.2825$\\
LaGd & $-$ & $0.30$ & $10$ & $0.18$ & $0.01$ & $0.2750$\\
LaGd & $-$ & $0.40$ & $15$ & $0.20$ & $0.01$ & $0.2670$\\
\hline
\hline
ThGd &$-$& $0.10$ & $5$ & $0.08$ & $0.01$ & $0.2090$ \\
ThGd &$-$& $0.20$ & $8$ & $0.09$ & $0.01$ & $0.1910$ \\
\hline
\end{tabular}
\caption{\label{table1}Parameter values.}
\end{center}
\end{table*}

The resulting phase diagrams also show the reentrant superconductivity, which is seen on Fig.~\ref{Fig1}a for $g \approx 0.72\,\sqrt{\text{eV}}$. As temperature $T$ is lowered, the system undergoes the first phase transition to a superconducting state ($P \rightarrow SC$) when $T = T_{\text{c}}$. After further cooling superconductivity is suppressed at $T = T_{SCF}$ and the alloy displays ferromagnetic properties until $T > T_{FD}$. For $T \leq T_{FD}$ superconductivity reappears in the $D$ phase, where it coexists with ferromagnetism. 

It is worth noting at this point that reentrant superconductivity due to the Kondo effect was observed for the first time in (La$_{1-x}$Ce$_{x})$Al$_{2}$ \cite{RibletRSC, MapleRSC}. These measurements showed destruction of superconductivity below the second critical temperature $T_{\text{c}2} < T_{\text{c}1}$, but the second phase transition, reintroducing superconductivity was not confirmed. This scenario with $T_{\text{c}1} = T_{\text{c}}$ and $T_{\text{c}2} = T_{SCF}$ is also present on the computed phase diagrams, e.g. on Fig.~\ref{Fig1}b for $g \geq 0.65\,\sqrt{\text{eV}}$. 
 
\begin{figure}
\begin{center}
\begin{tabular}{@{}c@{ }c@{ }c@{ }c@{}@{ }@{ }c@{ }c@{ }c@{ }c@{ }}
\multicolumn{1}{l}{\footnotesize \bf{a)}} \\[-0.9cm] 
    \includegraphics[width=7.5cm]{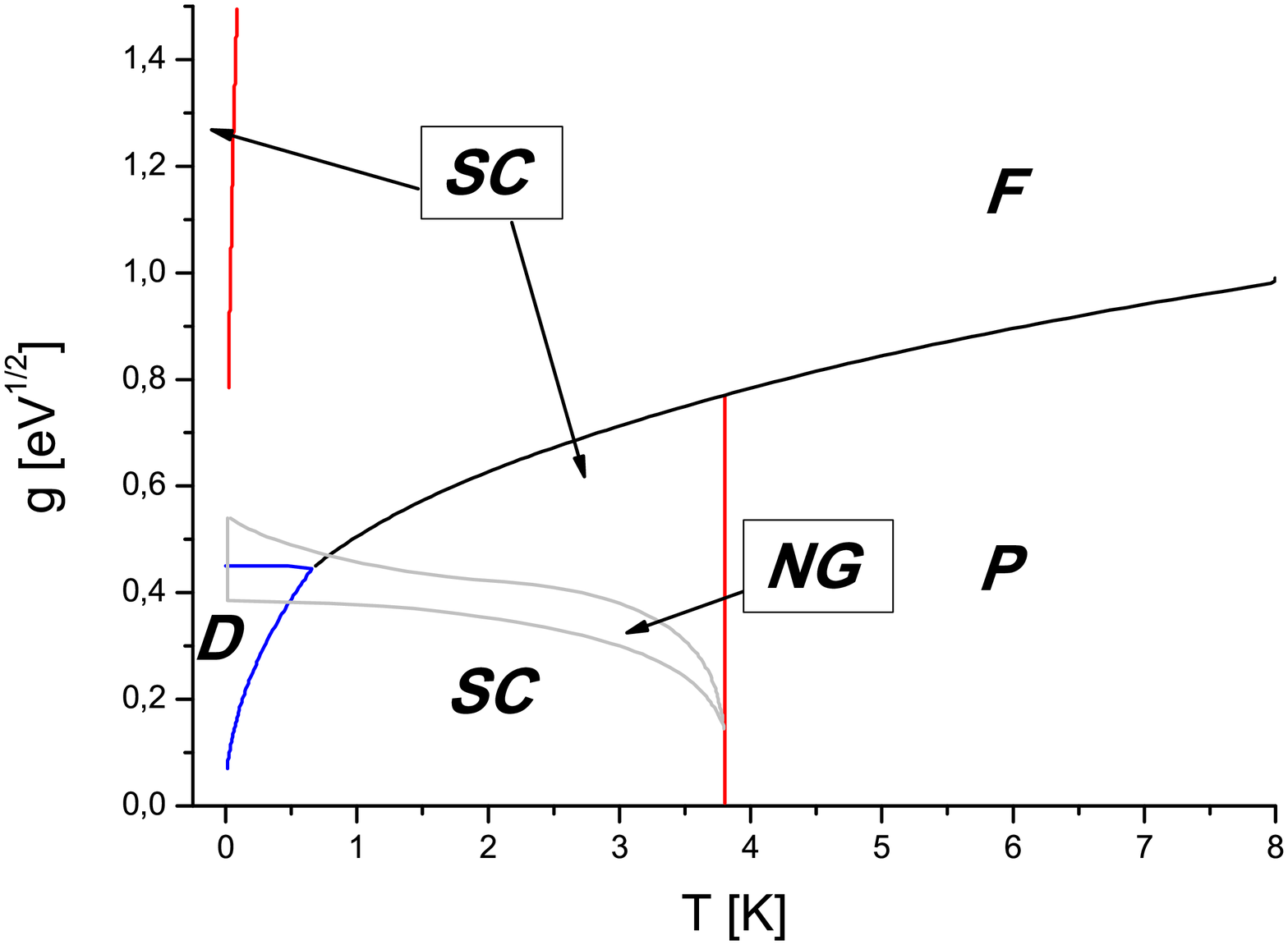} \\
\multicolumn{1}{l}{\footnotesize \bf{b)}} \\[-0.9cm]
    \includegraphics[width=7.5cm]{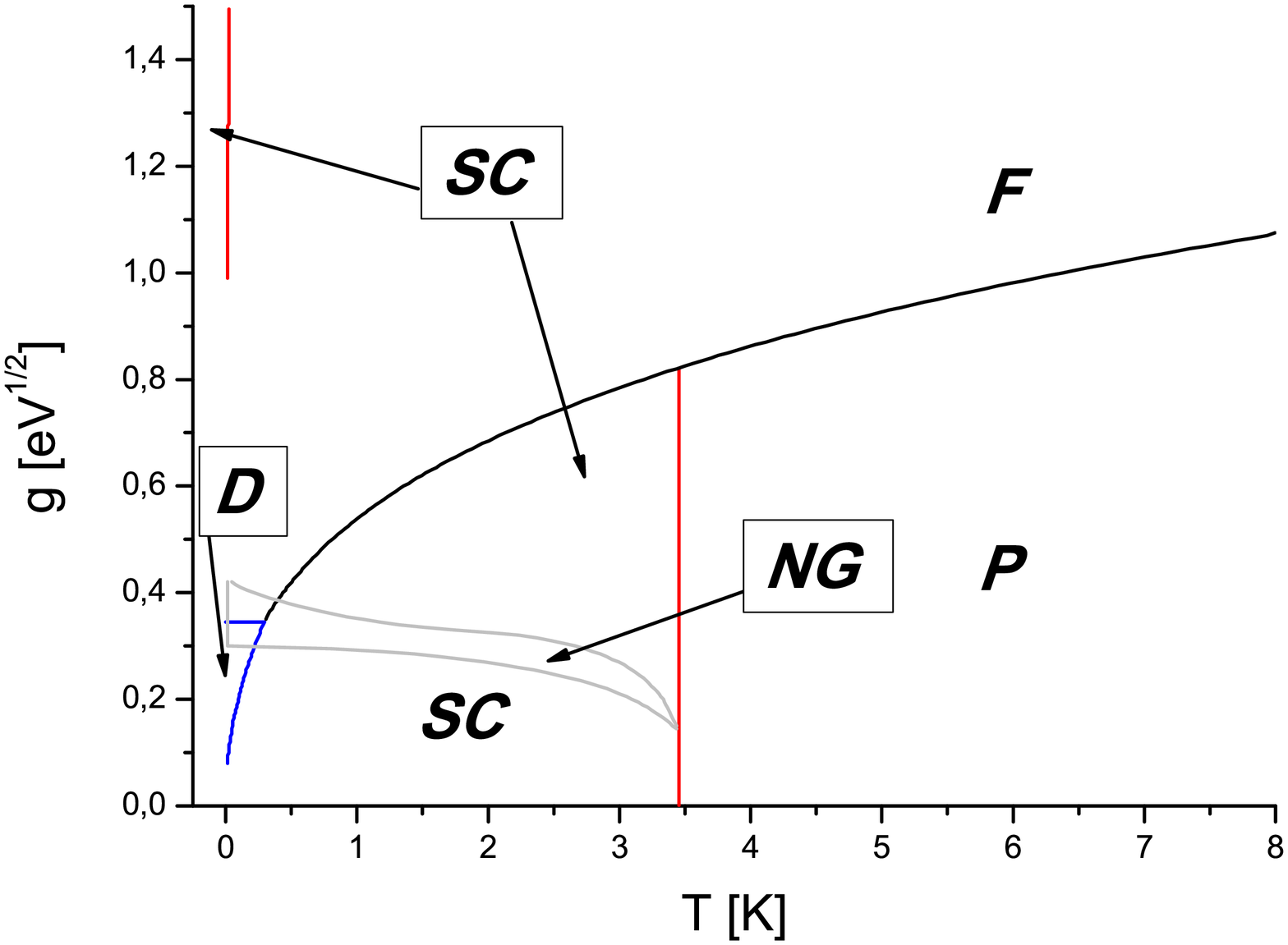} \\
\multicolumn{1}{l}{\footnotesize \bf{c)}}\\[-0.9cm]
\multicolumn{1}{l}{\includegraphics[width=7.5cm]{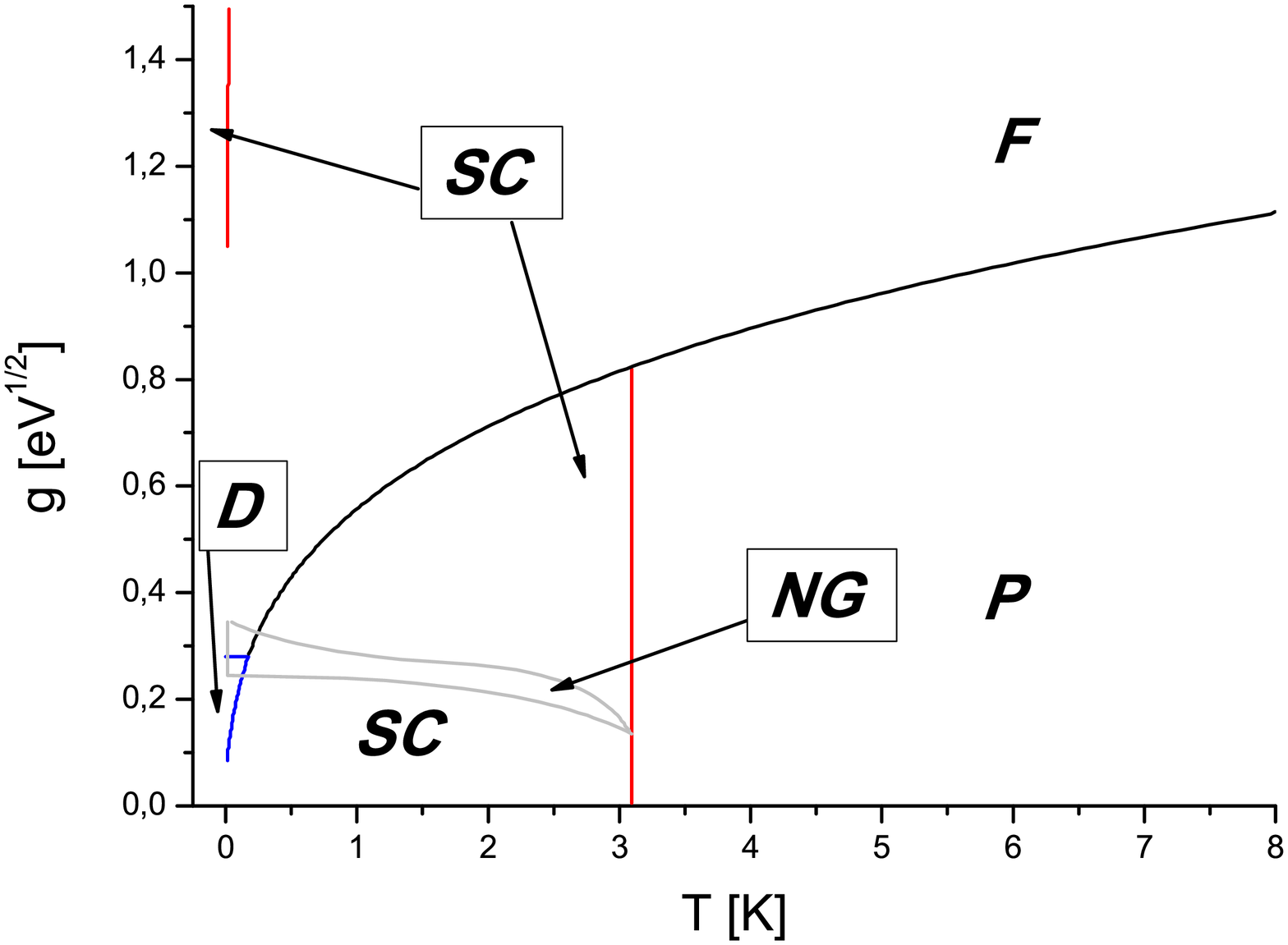}}\\
\end{tabular}
\caption{\label{Fig2}(Color online) Phase diagrams of LaGd for the values of $M$, $g$, $\delta$ and $G_{0}\rho_{F}$ given in Table~\ref{table1} and under varying Gd concentration : (a) $c = 0.10\ \%$ Gd, (b) $c = 0.20\ \%$ Gd and (c) $c = 0.30\ \%$ Gd.}
\end{center}
\end{figure}

\begin{figure}
\begin{center}
\begin{tabular}{@{}c@{ }c@{ }c@{ }c@{}@{ }@{ }c@{ }c@{ }c@{ }c@{ }}
\multicolumn{1}{l}{\footnotesize \bf{a)}} \\[-0.9cm]
    \includegraphics[width=7.5cm]{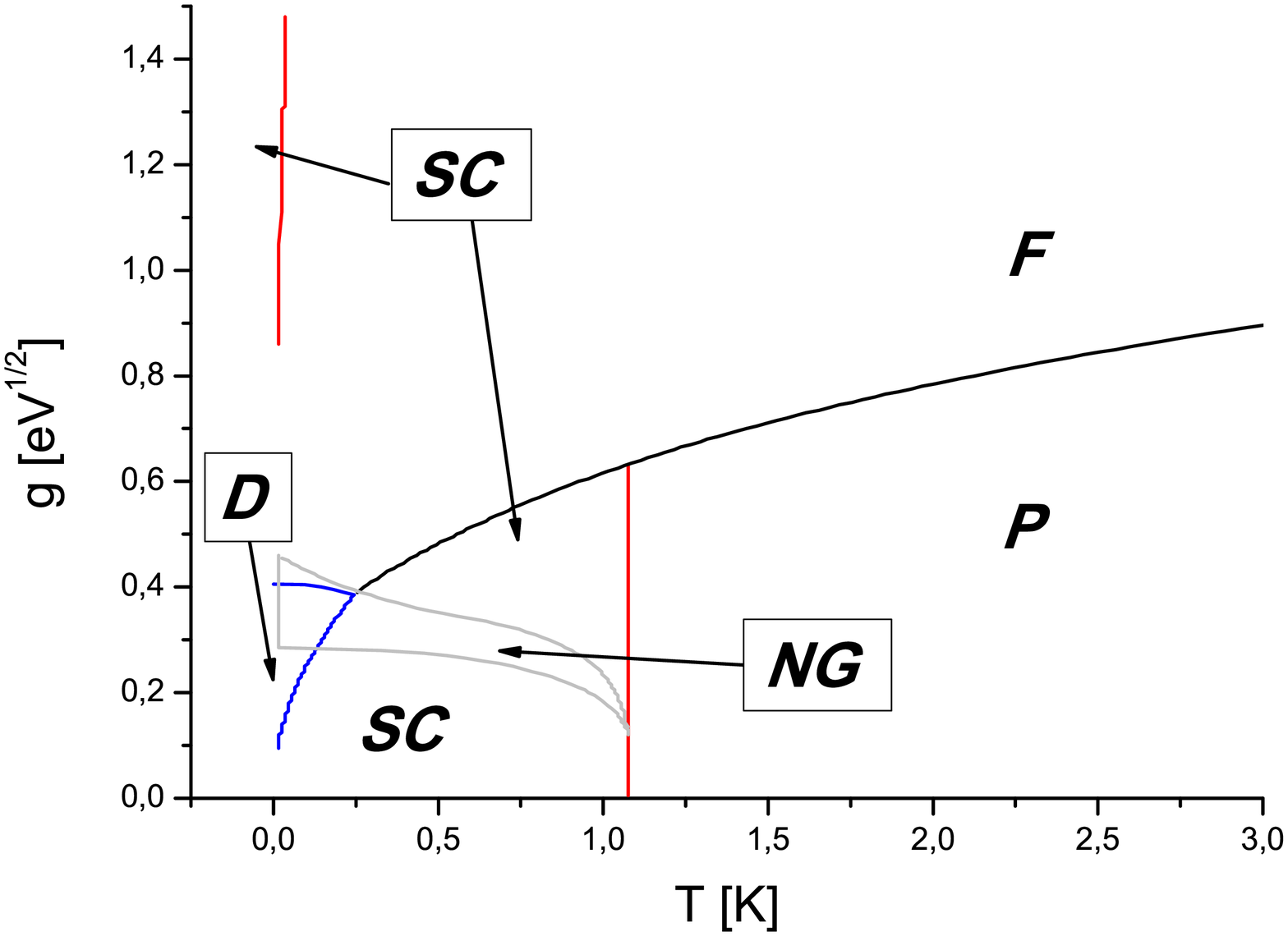}\\
\multicolumn{1}{l}{\footnotesize \bf{b)}}\\[-0.9cm] 
    \includegraphics[width=7.5cm]{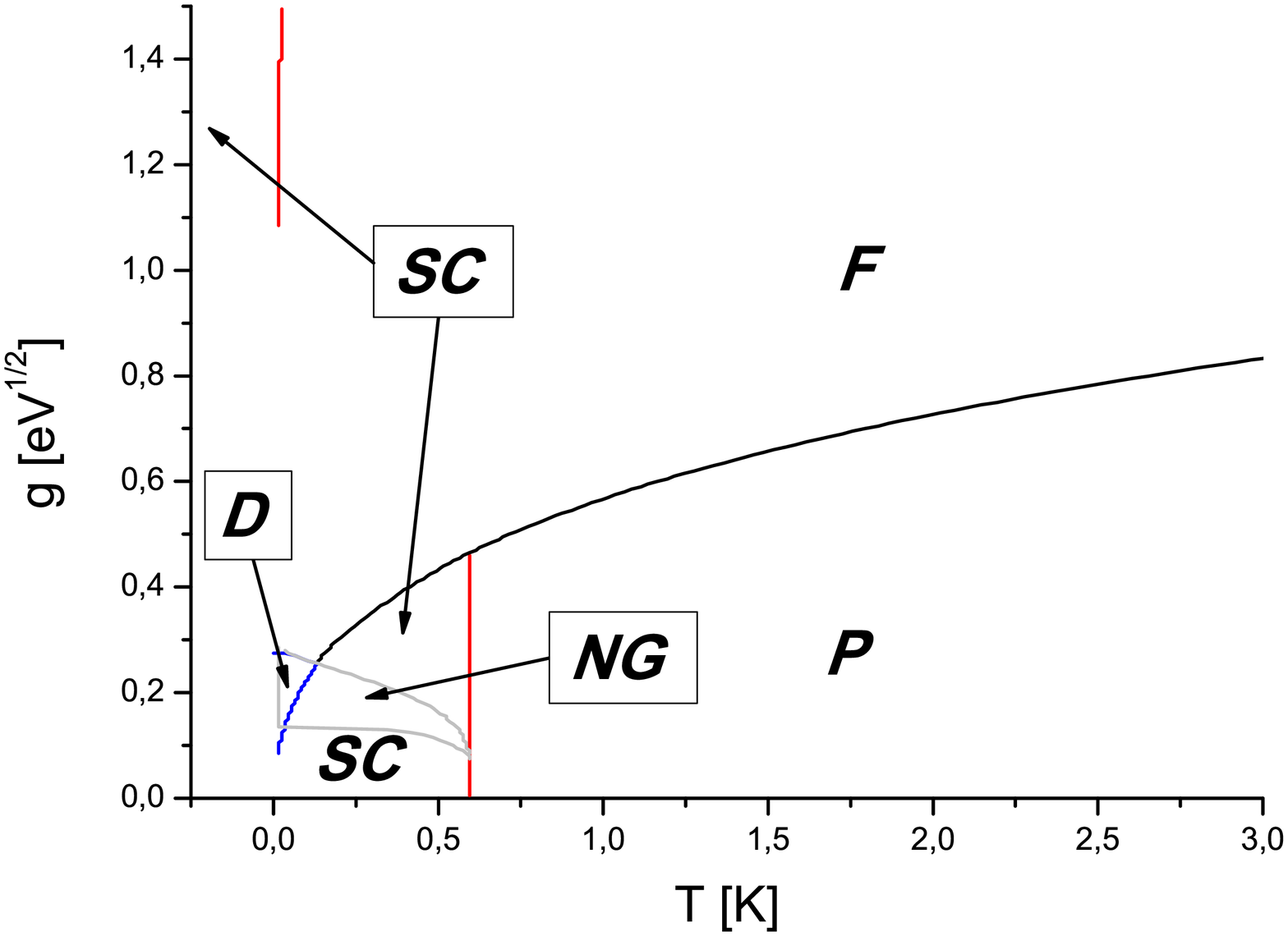}\\[-1em]
\end{tabular}
\caption{\label{Fig3}(Color online) Phase diagrams of ThGd for the values of $M$, $g$, $\delta$ and $G_{0}\rho_{F}$ given in Table~\ref{table1} and under varying Gd concentration : (a) $c = 0.10\ \%$ Gd, (b) $c = 0.20\ \%$ Gd.}
\end{center}
\end{figure}

The phase diagrams of LaGd and ThGd alloys, for which the free energy is given by Eq.~(\ref{F_1_2}) with $s = 7/2$, are depicted in Figs.~\ref{Fig2}~and~\ref{Fig3}. The values of parameters exploited in these computations are collected in Table~\ref{table1}. Figures~\ref{Fig2}~and~\ref{Fig3} show that the perturbative effect of Gd is larger than that of Ce impurities, viz., the area of $D$ phase is smaller than in Fig.~\ref{Fig1}. Furthermore, superconductivity is already suppressed at smaller values of $g$. This complies with experimental observations by Matthias et al., who showed that indeed the depression of superconductivity of doped lanthanum increases with the spin of the rare earth ions \cite{Matthias1}.

The $T_{SCF}$ temperature increases almost linearly with the value of $g$. One also observes decrease of $T_{FD}$ with $g$ and increase of $T_{PF}$ with $g$. Furthermore, rapid disappearance of $D$ phase is observed for critical values of $g$.

The intermediate phase $D$ in which superconductivity coexists with ferromagnetism of impurities is present on each phase diagram depicted in Figs.~\ref{Fig1},~\ref{Fig2} and \ref{Fig3}. The first experimental suggestions concerning the presence of coexistence phase in superconducting alloys was made by Matthias et al.\cite{Matthias1}. This hypothesis was later confirmed in Ce$_{1-x}$Gd$_{x}$Ru$_{2}$~\cite{MatthiasFerromagneticSC} and~Y$_{1-x}$Gd$_{x}$Os$_{2}$\cite{HS59}. However, the basic question, as to whether superconductivity and magnetic order occurs in the same volume element was still open. Wilhelm and Hillenbrand measured phase diagrams of Ce$_{1-x}$Tb$_{x}$Ru$_{2}$, which contain the coexistence phase\cite{MW70}. As temperature was lowered, one observed the following phase transitions, depending on the value of impurity concentration $c$:
\begin{enumerate}
\item A phase transition $P \rightarrow SC$, which occurs for sufficiently small $c$.
\item For intermediate values of $c$ the system undergoes a phase transition $P \rightarrow SC$ and then from $SC$ to $D$ phase.
\item Two phase transitions occur for sufficiently large values of $c$: $P \rightarrow F$ and then $F \rightarrow D$.
\item For some special value of $c$ the system undergoes a phase transition $P \rightarrow D$.  
\end{enumerate}  
The phase diagrams depicted in Figs.~\ref{Fig1},~\ref{Fig2}~and~\ref{Fig3} reveal the transitions 1--3. In particular, one phase transition $P \rightarrow SC$ (variant no. 1) is visible on every diagram depicted in Fig.~\ref{Fig1} for $g \leq 0.05\,\sqrt{\text{eV}}$, which states the weak coupling of Cooper pairs to impurity ions. The options 2 and 3 are clearly seen on every diagram. However, variant no. 3 is most distinct in Fig.~\ref{Fig1}a for $g \approx 0.9\,\sqrt{\text{eV}}$. The computed phase diagrams do not exhibit the fourth transition only, although, Fig.~\ref{Fig1}a shows that the phase transition $P \rightarrow D$ may appear for sufficiently large values of $g$, $G_{0}\rho_{\text{F}}$ and small $M$ and $c$, since in this case the $D$ region is relatively large. 

The specific heat measurements of Ce$_{1-x}$Tb$_{x}$Ru$_{2}$ revealed a short-range magnetic order in this compound, which is typical of spin-glasses\cite{MP71}. The coexistence of superconductivity with long-range ferromagnetic order was discovered a few years later, e.g. in (Er$_{1-x}$Ho$_{x}$)Rh$_{4}$B$_{4}$\cite{DJ78, OP85, BM85}. Phase diagrams of these compounds show the following phase transitions:
\begin{enumerate}
\item{$P \rightarrow SC$ for $x \approx 0.1$} 
\item{$P \rightarrow SC \rightarrow D$ for $x \approx 0.3$.}
\item{$P \rightarrow SC \rightarrow F \rightarrow D$ for $x \approx 0.4$.}
\item{$P \rightarrow F \rightarrow D$ for $x \approx 0.9$.}
\item{$P \rightarrow F$ for $x \approx 0.95$.}
\end{enumerate}
Theoretical phase diagrams depicted in Fig.~\ref{Fig1}--\ref{Fig3} do not reproduce the third scenario only.

The intermediate phase, where superconductivity coexists with long-range antiferromagnetic order was discovered in RMo$_{6}$Se$_{8}$ (R = Gd, Tb and Er)\cite{McCallumASC}, RRh$_{4}$B$_{4}$ (R = Nd, Sm and Tm)\cite{HamakerASC} and RMo$_{6}$S$_{8}$ (R = Gd, Tb, Dy and Er)\cite{IshikawaASC}. Upper critical field $H_{\text{c}2}$ in most of these compounds decreases below N\'{e}el temperature $T_{\text{N}}$. Accordingly, the superconducting state is perturbed by antiferromagnetically aligned impurity spins. However, a~rapid increase of $H_{\text{c}2}$ below $T_{\text{N}}$ was also observed, e.g. in SmRh$_{4}$B$_{4}$, GdMo$_{6}$S$_{8}$, TbMo$_{6}$S$_{8}$, Sn$_{1-x}$Eu$_{x}$Mo$_{6}$S$_{8}$, Pb$_{1-x}$Eu$_{x}$Mo$_{6}$S$_{8}$, La$_{1.2-x}$Eu$_{x}$Mo$_{6}$S$_{8}$ \cite{OF82, BM82}, implying enhancement of superconductivity by antiferromagnetism of impurities. 

Fischer et al. \cite{OF75} have pointed out that this boost of superconductivity may be due to the Jaccarino-Peter effect\cite{VJ62}, in which superconductivity is induced by applying a high magnetic field $\mathcal H$. This effect can arise in a type-II superconductor, in which the impurity magnetic moments are antiferromagnetically coupled to conduction electrons. This interaction generates an exchange field ${\mathcal H}_{\text{J}}$, which acts on the spins of conduction electrons equivalently to an applied magnetic field, viz., breaks the Copper pairs. However, the negative sign of the coupling between the magnetic moments and the conduction fermions spins, determines the direction of ${\mathcal H}_{\text{J}}$ to be opposite to that of $\mathcal H$. Thus, an applied magnetic field will be compensated by an exchange field, since the net magnetic field $\mathcal H_{\text{T}}$ is given by ${\mathcal H} - |{\mathcal H}_{\text{J}}|$. A given compound displays superconducting properties as long as the following relation holds:
\begin{equation}
\label{Eq:HPlimit}
-{\mathcal H}_{\text{p}} \leq {\mathcal H}_{\text{T}} \leq {\mathcal H}_{\text{p}}, 
\end{equation}    
where 
\begin{equation}
\label{Eq:Hp}
{\mathcal H}_{\text{p}} = \sqrt{\frac{\rho_{F}}{\chi_{\text{P}} - \chi_{\text{SC}}}}.
\end{equation}
$\chi_{\text{P}}$ and $\chi_{\text{SC}}$ denote the susceptibility of the normal and superconducting state respectively. ${\mathcal H}_{\text{p}}$, defined by Eq.~(\ref{Eq:Hp}) is the Chandrasekhar-Clogston limiting paramagnetic field \cite{BC62,AC62}.

The Jaccarino-Peter mechanism was observed experimentally in Eu$_{0.75}$Sn$_{0.25}$Mo$_{6}$S$_{7.2}$Se$_{0.8}$\cite{HM84}. In the low-temperature scale and with increasing applied magnetic field four phase transitions were recognized: $SC \rightarrow P \rightarrow SC \rightarrow P$. This effect can explain the reappearance of superconductivity for sufficiently large values of $g$ on theoretical phase diagrams (Figs.~\ref{Fig1},~\ref{Fig2}~and~\ref{Fig3}). For $T \approx 0.01 \text{K}$ it is firmly seen that the system undergoes the following transitions: $SC \rightarrow D \rightarrow F \rightarrow SC$, which can be interpreted as $SC \rightarrow P \rightarrow SC$, 
 since the given compound displays features characteristic of superconductors and magnetically ordered systems in a $D$ phase and in the $F$ phase superconductivity is suppressed.

In conclusion it is worth pointing out that the interplay between superconductivity and magnetism is believed to be a possible mechanism of high-$T_{\text{c}}$ superconductivity\cite{BK04}, since the undoped state of cuprate superconductors is a strongly insulating antiferromagnet. The existence of such a parent correlated insulator is viewed to be an essential feature of high temperature superconductivity.

\section{Critical temperature}
\label{Sec:CriticalTemperature}

According to section \ref{Sec:FreeEnergyMinimization}, Eq.~(\ref{Delta}) for the solution $\{\Delta \neq 0, \nu = 0\}$ reduces to the BCS gap equation
\begin{equation}
\label{DeltaBCS}
\Delta_{\text{BCS}} = \frac{1}{2} G_{0} \rho_{\text{F}} \int_{-\delta}^{\delta} \frac{\Delta_{\text{BCS}}}{E_{\text{BCS}}} \tanh \left(\tfrac{1}{2}\beta E_{\text{BCS}}\right) \text{d}\xi,
\end{equation}
\[
E_{\text{BCS}} = \sqrt{\xi^{2} + \Delta_{\text{BCS}}^{2}}.
\]
The transition temperature $T^{(\text{BCS})}_{\text{c}}$ in BCS theory, is defined as the boundary of the region beyond which there is no real, positive $\Delta_{\text{BCS}}$ satisfying (\ref{DeltaBCS}). Below $T^{(\text{BCS})}_{\text{c}}$ the solution $\Delta_{\text{BCS}} \neq 0$ minimizes the free energy and the system is in superconducting phase. Therefore, $T^{(\text{BCS})}_{\text{c}}$ can be obtained from Eq.~(\ref{DeltaBCS}) with $\Delta_{\text{BCS}} = 0$, which yields \cite{BCS}:
\begin{equation}
\label{TcBCS}
T^{(\text{BCS})}_{\text{c}} = 1.14 \delta \exp \bigl[-(G_{0} \rho_{\text{F}} )^{-1}\bigr].
\end{equation}
It should be possible to estimate the change in $T^{(\text{BCS})}_{\text{c}}$, since the density of states enters exponentially in (\ref{TcBCS}). However, significant deviations from Eq.~(\ref{TcBCS}) were observed experimentally for a number of superconductors containing magnetic impurities. This inadequacy of Eq.~(\ref{TcBCS}) is most distinct for large values of impurity concentration. BCS theory is therefore incapable to describe the superconducting alloys.

The results obtained in Sec.~\ref{Sec:PhaseDiagrams} show that phase transition to superconducting state in a superconductor with magnetic impurities, depending on the value of magnetic coupling constant, can be 1st or 2nd order. Next two subsections are concerned with computation of the transition temperature $T_{\text{c}}$ on the grounds of Eqs.~(\ref{Delta}), (\ref{nuEq}) and (\ref{F_1_2}). These calculations strongly depend on the order of the transition.
 
\subsection{Second order phase transition}
\label{Sec:SecondOrderPhaseTransition}
Expression for transition temperature $T_{\text{c}}$ for 2nd order phase transition can be computed analogously as in BCS theory. It suffices to put $\Delta = 0$ in Eqs.~(\ref{Delta}) and (\ref{nuEq}). Thus, one obtains the following set of equations for $T_{\text{c}} = 1/k \beta_{\text{c}}$,  
\begin{equation}
\label{DeltaC}
2 = G_{0} \rho_{\text{F}} \int_{-\delta}^{\delta} \frac{\text{d}\xi}{|\xi|} \frac{\sinh(\beta_{\text{c}} |\xi|)}{\cosh(\beta_{\text{c}} |\xi|) + \cosh(g \beta_{\text{c}} M f^{(s)}_{2}(\nu_{\text{c}}))},
\end{equation}
\begin{equation}
\label{NuC}
\nu_{\text{c}} = \frac{c g}{M} \frac{\sinh(g \beta_{\text{c}} M f^{(s)}_{2}(\nu_{\text{c}}))}{\cosh(g \beta_{\text{c}} M f^{(s)}_{2}(\nu_{\text{c}})) + \cosh(\beta_{\text{c}} |\xi|)} + f^{(s)}_{2}(\nu_{\text{c}}),
\end{equation}
where $\nu_{\text{c}} = \nu(\beta_{\text{c}})$, $s = 1/2,\,7/2$. Numerical analysis show that in the low-temperature scale $\nu_{\text{c}}$ is almost independent in $T$, viz.,
\begin{itemize}
\item {for spin $1/2$ impurities:}
\[
\nu_{c} \approx \left\{ \begin{array}{rl}
0, & \text{for } x \leq 0.0010, \\
\nu(0) = \frac{c g}{M}, & \text{for } x \geq 0.0019,
\end{array} \right.
\]
\item {for spin $7/2$ impurities:}
\[
\nu_{\text{c}} \approx \nu(0) = \frac{c g}{M}.
\]
\end{itemize}
This result complies with experimental data, showing that the perturbative effect of impurities on superconductivity is an increasing function of their spin. Furthermore, the reduction of $T_{\text{c}}$ by adding a small amount of spin $1/2$ impurities can be satisfactorily described by Eq.~(\ref{TcBCS}) by adjusting the value of $G_{0} \rho_{\text{F}}$, since for $\nu_{\text{c}} = 0$ Eqs.~(\ref{DeltaC}) and (\ref{NuC}) reduce to $T_{\text{c}}^{(\text{BCS})}$. Accordingly, the set of Eqs.~(\ref{DeltaC}), (\ref{NuC}) are solved under the assumption that $\nu_{c}=\frac{cg}{M}$.

\begin{figure}[b]
\begin{center}
\includegraphics[width=6.5cm]{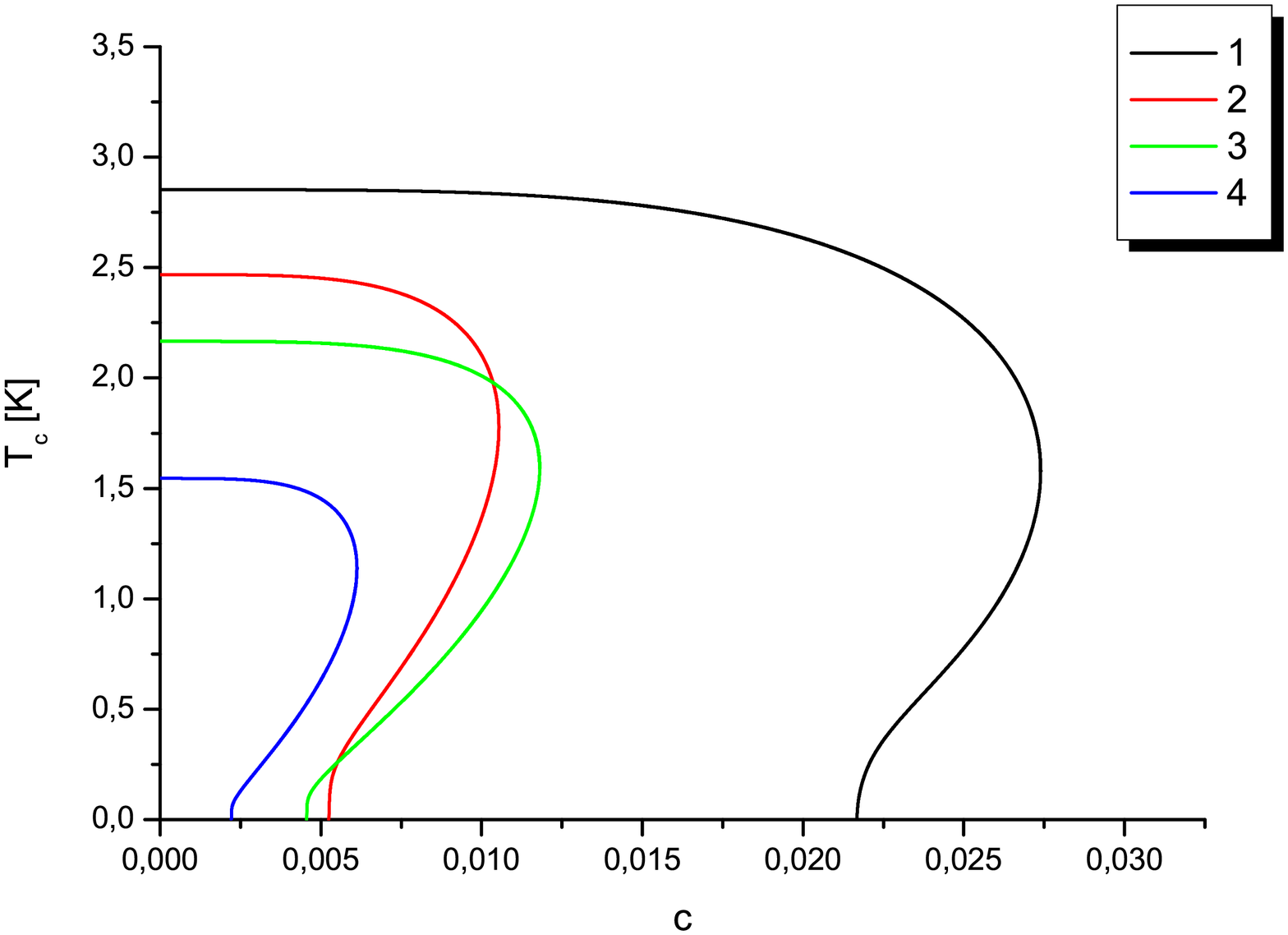}
\end{center}
\caption{\label{Fig4}
(Color online) The superconducting transition temperature $T_{\text{c}}$ under varying impurity concentration $c$ for superconducting alloy, containing $1/2$ impurities. The values of the parameters $g$, $G_{0} \rho_{\text{F}}$ and $M$ are given in table \ref{tableTc}.}
\end{figure}
\begin{figure}[b]
\begin{center}
\includegraphics[width=6.5cm]{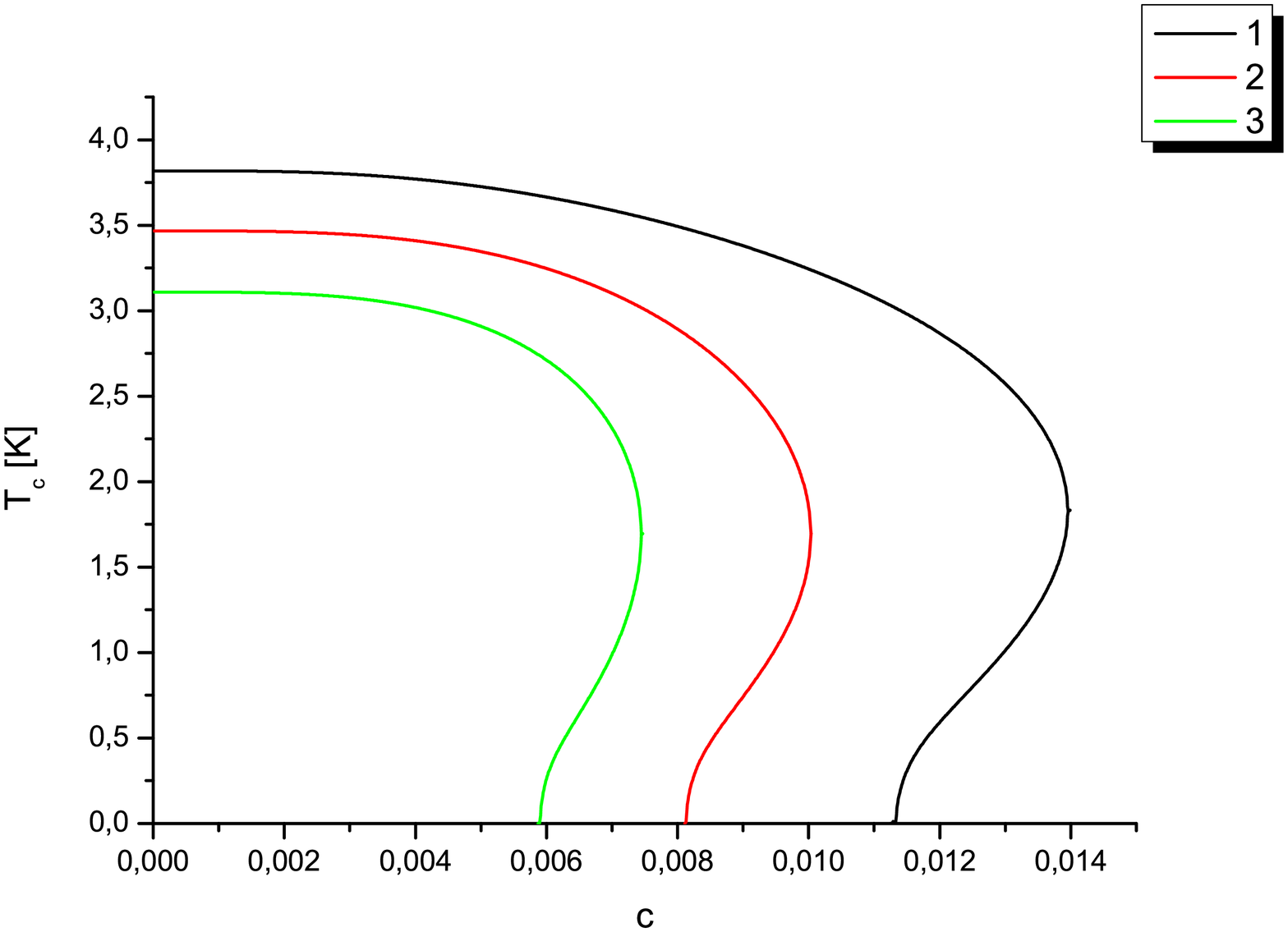}
\end{center}
\caption{\label{Fig5}
(Color online) The superconducting transition temperature $T_{\text{c}}$ under varying impurity concentration $c$ for superconducting alloy, containing $7/2$ impurities. The values of the parameters $g$, $G_{0} \rho_{\text{F}}$ and $M$ are given in table \ref{tableTc}.}
\end{figure}

\begin{table*}
\tabcolsep12pt
\begin{center}
\begin{tabular}{*{7}{|c}|}
\hline Impurity spin & Curve no. & $M$ & $g\ [\sqrt{\text{eV}}]$ & $\delta\ [\text{eV}]$ & $G_{0} \rho_{F}$ \\
\hline\hline
\multirow{4}{*}{$1/2$} & $1$ & $1$ & $0.10$ & $0.01$ & $0.2610$ \\
 & $2$ & $4$ & $0.189$ & $0.01$ & $0.2515$ \\
 & $3$ & $7$ & $0.19$ & $0.01$ & $0.2435$ \\
 & $4$ & $8$ & $0.23$ & $0.01$ & $0.2250$\\
\hline\hline
\multirow{3}{*}{$7/2$} & $1$ & $5$ & $0.16$ & $0.01$ & $0.2825$ \\
 & $2$ & $10$ & $0.18$ & $0.01$ & $0.2750$ \\
 & $3$ & $15$ & $0.20$ & $0.01$ & $0.2670$ \\
\hline
\end{tabular}
\caption{\label{tableTc}Parameter values.}
\end{center}
\end{table*}

The solution for $T_{\text{c}}$ resulting from Eqs.~(\ref{DeltaC}), (\ref{NuC}) for $s = 1/2,\,7/2$ and under varying impurity concentration $c$ is depicted in Figs.~\ref{Fig4},~\ref{Fig5}. In general, the values of the parameters $M$, $g$, $\delta$ and $G_{0} \rho_{\text{F}}$, depend on the impurity concentration $c$. However, to minimize the number of adjustable parameters, the values of $M$, $g$, $\delta$ and $G_{0} \rho_{\text{F}}$ have been kept constant.

Figures~\ref{Fig4},~\ref{Fig5} show how variation of $c$ affects $T_{\text{c}}(c)$. For certain values of $c$ (e.g. $c \in (0.0225, 0.0275)$ for curve no.~1 on Fig.~\ref{Fig4}) Eq.~(\ref{DeltaC}) with $\nu_{\text{c}} \approx \nu(0)$ possesses two solutions: $T_{\text{c}1}$, $T_{\text{c}2}$. Thus, superconductivity is suppressed below $T_{\text{c}2}$. The dependence of $T_{\text{c}}$ on $c$ shown in Figs.~\ref{Fig4},~\ref{Fig5} does not indicate the existence of a third transition temperature $T_{\text{c}3}$. This, can be a consequence  of the applied approximation in which $M$, $g$, $\delta$ i $G_{0} \rho_{\text{F}}$ do not depend on $c$. However, the dependence of $T_{\text{c}}$ on impurity concentration $c$ shown in Fig.~\ref{Fig4} is typical of the La$_{1-x}$Ce$_{x}$Al$_{2}$ alloy\cite{MapleConvAndUnconvSC}. 

The destructive influence of impurities on superconductivity is analogous to the effect of an external magnetic field $\mathcal H$ applied to a superconducting compound. Sarma \cite{Sarma63} obtained a numerical solution for $T_{\text{c}}({\mathcal H})$ of the system described by a Hamiltonian $H_{\text{S}} = H_{\text{BCS}} + \mu_{\text{B}} {\mathcal H} \sigma_{z}$. His result for $T_{\text{c}}({\mathcal H})$ agrees qualitatively with $T_{\text{c}}(c)$ graphs depicted in Figs.~\ref{Fig4}~and~\ref{Fig5}, since the expression for $T_{\text{c}}({\mathcal H})$ obtained in Ref.~\onlinecite{Sarma63} is of the similar form to Eq.~(\ref{DeltaC}) with $g M f^{(s)}(\nu_{\text{c}})$ replacing $\mu_{\text{B}} {\mathcal H}$.

\subsection{The first order phase transition}

In the case of 1st order phase transition, the assumption that the gap parameter $\Delta$ vanishes at the transition temperature does not hold. According to the results obtained in the previous section, one can expect that $T_{\text{c}}$ possesses three solutions ($T_{\text{c}1} \leq T_{\text{c}2} \leq T_{\text{c}3}$) for certain values of $g$ and $c$. These solutions can be determined numerically from the following equations
\begin{equation}
\label{Tc_1}
T_{\text{c}1} : \qquad F_{P} - F_{SC} = 0,
\end{equation} 
\begin{equation}
\label{Tc_2}
T_{\text{c}2} : \qquad F_{SC} - F_{\Phi} = 0, \qquad \Phi = D, F,
\end{equation} 
\begin{equation}
\label{Tc_3}
T_{\text{c}3} : \qquad F_{F} - F_{D} = 0. 
\end{equation} 

The existence of $T_{\text{c}3}$ depends on the type of phase transition occuring at $T_{\text{c}2}$. If the system undergoes a phase transition to ferromagnetic phase at $T_{\text{c}2}$, then $T_{\text{c}3} > 0$ for certain values of $g$ (Sec.~\ref{Sec:PhaseDiagrams}). If $T_{\text{c}2} = T_{SCD}$, then $T_{\text{c}3} = 0$ and the system does not reenter the superconducting phase ($SC$ or $D$).

\begin{figure}
\begin{center}
\includegraphics[height=6.5cm]{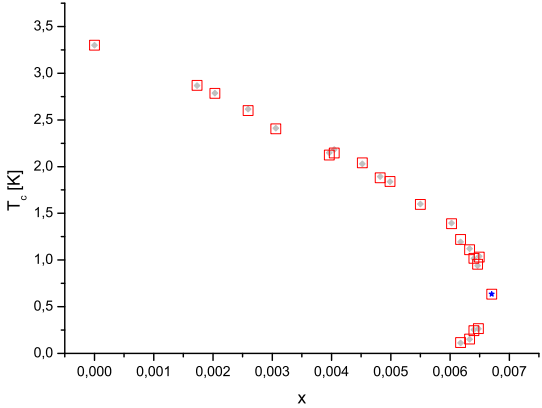}
\end{center}
\caption{\label{Fig6}
(Color online) The superconducting transition temperature vs Ce impurity concentration $x$ for (La$_{1-x}$Ce$_{x}$)Al$_{2}$. Squares denote theoretical values of $T_{\text{c}}$ obtained from Eqs.~(\ref{Tc_1})--(\ref{Tc_3}). Rhombs are experimental points from Ref.~\onlinecite{MapleRSC}. The star at the point $x= 0.0067$ denotes the experimentally estimated turning point in $T_{\text{c}}$, above which superconductivity is destroyed.}
\end{figure}

\begin{figure}
\begin{center}
\includegraphics[height=6.5cm]{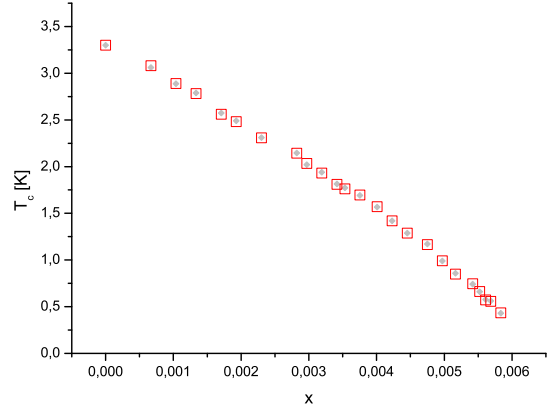}
\end{center}
\caption{\label{Fig7}
(Color online) The superconducting transition temperature vs Gd impurity concentration $x$ for (La$_{1-x}$Gd$_{x}$)Al$_{2}$. Squares denote theoretical values of $T_{\text{c}}$ obtained from Eqs.~(\ref{Tc_1})--(\ref{Tc_3}). Rhombs are experimental points from Ref.~\onlinecite{MapleLaGdAl2}. The value of critical concentration $x_{\text{cr}}$, where $T_{\text{c}} = 0$ is $0.59$ at. \% Gd.}
\end{figure}

\begin{figure*}
\begin{center}
\includegraphics[width=12.5cm]{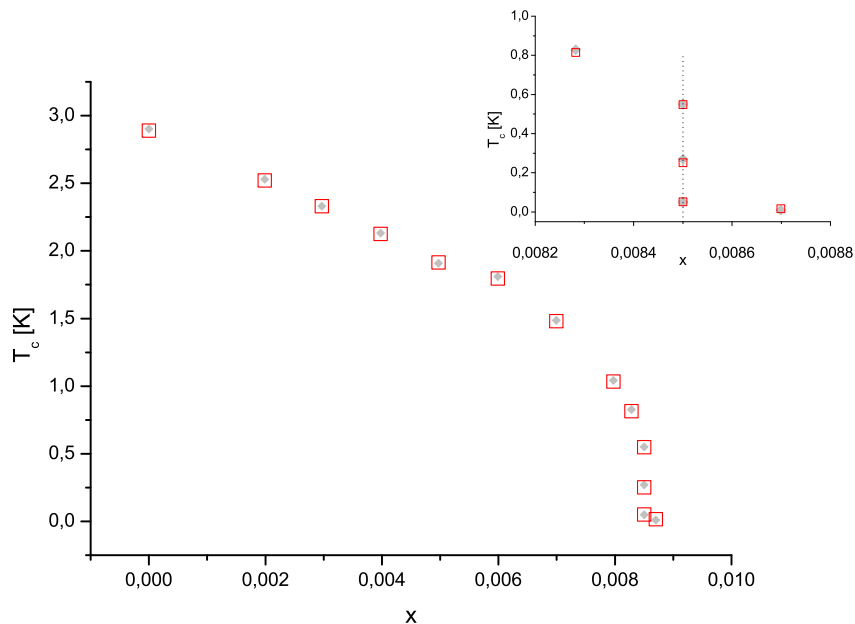}
\end{center}
\caption{\label{Fig8}
(Color online) The superconducting transition temperature vs Ce impurity concentration $x$ for (La$_{0.8-x}$Y$_{0.20}$)Ce$_{x}$. Squares denote theoretical values of $T_{\text{c}}$ obtained from Eqs.~(\ref{Tc_1})--(\ref{Tc_3}). Rhombs are experimental points from Ref.~\onlinecite{Winzer}. For $x = 0.0085$ there are three $T_{\text{c}}$ solutions, which demonstrate reentrant superconductivity observed experimentally in (La$_{0.8-x}$Y$_{0.20}$)Ce$_{x}$ at this value of $x$ (inset).}
\end{figure*}

In order to obtain a reliable comparison of the transition temperature resulting from Eqs.~(\ref{Tc_1})--(\ref{Tc_3}) to experimental data, we let the parameters $G_{0} \rho_{\text{F}},$ $g$ and $M$ vary with impurity concentration. The parameters are then adjusted independently for each experimental point. The best fitting of $T_{\text{c}}(c)$ to experiment for La$_{1-x}$Ce$_{x}$Al$_{2}$, La$_{1-x}$Gd$_{x}$Al$_{2}$ and (La$_{0.8-x}$Y$_{0.20}$)Ce$_{x}$ is plotted in Figs.~\ref{Fig6}, \ref{Fig7}~and~\ref{Fig8}. In all cases very good quantitative agreement with experiment was found.

Fig. \ref{Fig6} shows two phase transitions for $x~\in~(0.0064,\,0.0067)$. The first phase transition ($P\rightarrow SC$) occurs at $T_{\text{c}1}$ and the second transition (back to normal state) appears at $T_{\text{c}2}$. The reentrant superconductivity (the transitions $SC \rightarrow P \rightarrow SC$) is clearly visible in (La$_{0.8-x}$Y$_{0.20}$)Ce$_{x}$ for $x = 0.0085$ (Fig.~\ref{Fig8}).

The theoretical curve $T_{\text{c}}(x)$ resulting from AG theory also yields very good quantitative agreement with experimental data for La$_{1-x}$Gd$_{x}$Al$_{2}$. However, significant deviations from this theory were observed in (La$_{1-x}$Ce$_{x}$)Al$_{2}$ and (La$_{0.8-x}$Y$_{0.20}$)Ce$_{x}$, since AG predicts only single-valued solutions for the superconducting transition temperature\cite{MapleConvAndUnconvSC}. The model studied here gives considerably better results for $T_{\text{c}}$ than previous theories, although analytic expression for $T_{\text{c}}$ cannot be obtained for 1st order phase transitions.

\section{Concluding remarks}

We have shown that the thermodynamics of a BCS superconductor, perturbed by a
reduced s-d interaction is solvable if the impurity density
is sufficiently small. The essential properties of dilute magnetic superconducting alloys have been demonstrated: the decrease of superconducting transition temperature $T_{\text{c}}$ with increasing impurity concentration, the existence of reentrant and gapless superconductivity and the presence of Jaccarino-Peter compensation related to the magnetic field induced superconductivity. Good quantitative agreement of the resulting dependence of $T_{\text{c}}$ on impurity concentration was demonstrated for several superconducting compounds: La$_{1-x}$Ce$_{x}$Al$_{2}$, La$_{1-x}$Gd$_{x}$Al$_{2}$ and (La$_{0.8-x}$Y$_{0.20}$)Ce$_{x}$. The theory presented provides better agreement with experiment than earlier AG, MHZ theories and their various extensions and refinements given in Refs.~\onlinecite{PS75a, PS75b, TM76, MHZ76, MJ90}.  

We have also performed a fit of specific-heat and critical field curves for La$_{1-x}$Ce$_{x}$Al$_{2}$, LaCe, LaGd, and ThGd. We shall report on this study elsewhere \cite{DB10p}. 

These investigations will be extended to include the effect
of a general s-d exchange interaction $V_{\text{s-d}}$ and a BCS-type attraction between Cooper pairs $V_{\text{4f}}$\cite{TB, DB08, DB09} in order to study other properties of superconducting alloys. 

\bibliography{2fermion}
\end{document}